\documentclass[aps,prd,twocolumn,preprintnumbers,superscriptaddress,nofootinbib,floatfix]{revtex4-1}

\usepackage[dvipsnames]{xcolor}
\usepackage{amssymb,amsmath,amsfonts}
\usepackage{bm}
\usepackage{graphicx}
\usepackage{epstopdf}
\usepackage{hyperref}
\usepackage{array}
\usepackage{colortbl}
\usepackage[utf8]{inputenc}
\usepackage{soul}
\usepackage{color}
\usepackage[T1]{fontenc}

\enlargethispage{\baselineskip}

\definecolor{steelblue}{RGB}{25,25,112}
\definecolor{dullblue}{rgb}{0,0.298,0.49}
\definecolor{darkred}{rgb}{0.545,0,0}
\definecolor{blue2}{cmyk}{1, 0.1, 0.1, 0}

\hypersetup{colorlinks,linkcolor={darkred},citecolor={dullblue},urlcolor={dullblue}}  

\everymath{\displaystyle}

\newcommand{\MC}{M_{\rm AMC}}
\newcommand{\RC}{R_{\rm AMC}}
\newcommand{\rC}{\rho_{\rm AMC}}
\newcommand{\nC}{n_{\rm AMC}}
\newcommand{\fC}{f_{\rm AMC}}
\newcommand{\NC}{N_{\rm AMC}}

\renewcommand\({\left(}
\renewcommand\){\right)}
\renewcommand\[{\left[}
\renewcommand\]{\right]}

\begin{document}

\title{Stellar Disruption of Axion Miniclusters in the Milky Way}

\newcommand{\GRAPPA}{\affiliation{Gravitation Astroparticle Physics Amsterdam (GRAPPA), Institute for Theoretical Physics Amsterdam and Delta Institute for Theoretical Physics, University of Amsterdam, Science Park 904, 1098 XH Amsterdam, The Netherlands}}
\newcommand{\OKC}{\affiliation{The Oskar Klein Centre for Cosmoparticle Physics,
    AlbaNova University Center,\\
	Roslagstullsbacken 21,
	SE--106\.91 Stockholm,
	Sweden}} 
\newcommand{\IFCA}{\affiliation{Instituto de F\'isica de Cantabria (IFCA, UC-CSIC), Av.~de
Los Castros s/n, 39005 Santander, Spain}}
\newcommand{\INFN}{\affiliation{INFN, Laboratori Nazionali di Frascati, C.P. 13, 100044 Frascati, Italy}}

\author{Bradley J. Kavanagh}\email[Electronic address: ]{kavanagh@ifca.unican.es} \IFCA \GRAPPA
\author{Thomas D. P. Edwards}\email[Electronic address: ]{thomas.edwards@fysik.su.se} \OKC \GRAPPA
\author{Luca Visinelli}\email[Electronic address: ]{luca.visinelli@sjtu.edu.cn} \GRAPPA \INFN
\author{Christoph Weniger}\email[Electronic address: ]{c.weniger@uva.nl} \GRAPPA

\date{\today}

\begin{abstract}
    Axion miniclusters are dense bound structures of dark matter axions that are predicted to form in the post-inflationary Peccei-Quinn symmetry breaking scenario. Although dense, miniclusters can easily be perturbed or even become unbound by interactions with baryonic objects such as stars. Here, we characterize the spatial distribution and properties of miniclusters in the Milky Way (MW) today after undergoing these stellar interactions throughout their lifetime. We do this by performing a suite of Monte Carlo simulations which track the miniclusters' structure and, in particular, accounts for partial disruption and mass loss through successive interactions. 
    We consider two density profiles --- Navarro-Frenk-White (NFW) and Power-law (PL) --- for the individual miniclusters in order to bracket the uncertainties on the minicluster population today due to their uncertain formation history.
    For our fiducial analysis at the Solar position, we find a survival probability of 99\% for miniclusters with PL profiles and 46\% for those with NFW profiles. Our work extends previous estimates of this local survival probability to the entire MW. We find that towards the Galactic center, the survival probabilities drop drastically. Although we present results for a particular initial halo mass function, our simulations can be easily recast to different models using the provided data and code (\href{https://github.com/bradkav/axion-miniclusters/}{github.com/bradkav/axion-miniclusters}).
    Finally, we comment on the impact of our results on lensing, direct, and indirect detection.

\end{abstract}

\maketitle

\section{Introduction}
\label{sec:intro}

Both the dark matter (DM) and Strong-CP problems can be solved by introducing a new global symmetry into the Standard Model (SM) of particle physics~\cite{Peccei:1977hh, Peccei:1977ur}. This new Peccei-Quinn (PQ) symmetry $U_{\rm PQ}(1)$ predicts a hypothetical particle known as the QCD axion~\cite{Weinberg:1977ma, Wilczek:1977pj}. Unlike weakly interacting massive particles, DM axions are typically much lighter than the rest of the SM~\cite{Abbott:1982af, Dine:1982ah, Preskill:1982cy}. Their production mechanism must therefore rely upon non-thermal processes to ensure they are non-relativistic at the time of recombination. These non-thermal processes generically produce gravitationally bound \textit{clumps} of axions known as axion miniclusters. In this paper, we characterize the degree to which tidal interactions can change the properties of these miniclusters over the lifetime of the Milky Way (MW). 

\vskip 3pt

Worldwide, there is an active research program searching for QCD axion DM, as well as more general axion-like particles~\cite{Arvanitaki:2009fg}. Unfortunately, the axion's coupling to SM particles is expected to be extremely small and therefore challenging to probe. The majority of searches rely upon modifications to Maxwell's equations due to the axion-photon coupling $g_{a\gamma\gamma}$~\cite{Sikivie:1983ip, Sikivie:1985yu, Wilczek:1987mv, Krasnikov:1996bm, Li:2009tca, Visinelli:2013fia, Tercas:2018gxv, Visinelli:2018zif}. This has inspired a number of terrestrial direct search strategies~\cite{Hagmann:1998cb, Asztalos:2001tf, Asztalos:2003px, Graham:2015ouw, Braine:2019fqb, Kenany:2016tta, Brubaker:2016ktl, TheMADMAXWorkingGroup:2016hpc, Majorovits:2016yvk, Alesini:2017ifp,  Alesini:2019nzq, Kahn:2016aff, Ouellet:2018beu, Budker:2013hfa, Barbieri:2016vwg, Lee:2020cfj}.\footnote{For additional details, see the recent reviews on axion cosmology~\cite{Kawasaki:2013ae, Marsh:2015xka}, models of QCD axions~\cite{DiLuzio:2020wdo}, and detection techniques~\cite{Irastorza:2018dyq, Sikivie:2020zpn}.}
Indirect probes of axions also utilize the axion-photon coupling but instead in astrophysical settings. For example, Galactic axions can convert into radio photons in the magnetic field of a neutron star (NS)~\cite{Pshirkov:2007st, Huang:2018lxq, Hook:2018iia, Safdi:2018oeu, Leroy:2019ghm}. If the NS is locked in a binary system with an intermediate mass black hole, it may be possible to uniquely detect the radio signal jointly with a characteristic gravitational wave signature~\cite{Edwards:2019tzf}. Substructures in the axion distribution may have dramatic effects on all such searches.

The production of QCD axion DM is tightly connected to the thermal history of the Universe. 
After the $U_{\rm PQ}(1)$ symmetry is spontaneously broken, the axion field relaxes towards the bottom of its potential. 
When the Universe has cooled down to the QCD phase transition, non-perturbative QCD instantons lead to the explicit breaking of the PQ symmetry~\cite{Callan:1977gz, Gross:1980br}, giving rise to a new CP-conserving minimum in the potential.
After the QCD phase transition, the axion field undergoes coherent oscillations about this minimum, damped by the Hubble expansion rate $H$. This process is known as the vacuum realignment mechanism~\cite{Abbott:1982af, Preskill:1982cy, Dine:1982ah}. The DM energy density in this scenario is stored in the coherent oscillations of the axion condensate and depends on the initial value of the axion field when the PQ symmetry breaks,
which is parametrized by the initial misalignment angle $\theta_i$.
We generally expect the axion energy density to be proportional to $\theta_i^2$ except around $\theta_i \sim \pi$ where the non-harmonic terms in the axion potential become important~\cite{Strobl:1994wk, Bae:2008ue, Hertzberg:2008wr, Visinelli:2009kt, Visinelli:2014twa}.
The properties of the axion condensate crucially depend on whether the spontaneous breaking of the $U_{\rm PQ}(1)$ symmetry occurs before or after the end of inflation.
In the pre-inflationary scenario, the value of $\theta_i$ is uniquely selected over the whole observable Universe. 

\vskip 3pt

Here, we consider the opposite scenario in which the PQ symmetry is broken \textit{after} the end of inflation --- the post-inflationary scenario. In this case, the initial misalignment angle $\theta_i$ takes different values in different patches of the observable universe, since no patch has been selected by the inflationary process. In this post-inflationary scenario, self-gravitating substructures called axion miniclusters (AMCs)~\cite{Hogan:1988mp, Kolb:1993hw, Kolb:1993zz} are expected to form. Moderate $\mathcal{O}\(1\)$ overdensities initially lead to the formation of minicluster `seeds'~\cite{Vaquero:2018tib} which later collapse into gravitationally bound AMCs at around matter-radiation equality~\cite{Hogan:1988mp, Kolb:1993zz, Kolb:1993hw, Zurek:2006sy}.\footnote{Note that we will use the terms miniclusters and AMCs interchangeably throughout the paper.}
Instead, AMCs cannot form in the pre-inflationary scenario even when the initial conditions of the QCD axion field are extremely fine-tuned~\cite{Fukunaga:2020mvq}. 

Significant progress has been made towards solving the early universe dynamics of the axion; numerical simulations loosely constrain the fraction of cold DM axions in these bound structures $f_{\rm AMC}$ to be $\mathcal{O}(1\%-100\%)$~\cite{Vaquero:2018tib, Buschmann:2019icd}.
This fraction $f_{\rm AMC}$ plays a fundamental role in the prospects for axion DM detection. For example, if most of the DM is bound in AMCs,
the probability of having a substantial DM density near Earth
may drop drastically, due to the rarity of Earth-AMC encounters~\cite{Sikivie:2006ni}, making direct detection methods ineffective.
Similarly, the encounter rate of AMCs with a single NS is likely to be low, rendering radio observations of individual NSs ineffective in the search for axion-photon conversion.
Fortunately, encounters between miniclusters and the MW \textit{population} of NSs can give rise to interesting transient radio signals, as we show in our companion paper Ref.~\cite{Edwards:2020afl}.

It is possible to assess the fraction of cold axions bound in AMCs through femto-lensing induced by individual miniclusters~\cite{Kolb:1995bu, Katz:2018zrn}, micro-lensing from minicluster halos formed after matter-radiation equality from hierarchical merging~\cite{Fairbairn:2017dmf, Fairbairn:2017sil, Ellis:2020gtq}, and minicluster lensing of highly magnified stars~\cite{Dai:2019lud}. 
These studies typically treat $f_\mathrm{AMC}$ as a constant, but miniclusters interacting with their environment in fact cause $f_{\rm AMC}$ to become dependent on both time and spatial position. Tidal interactions of miniclusters with larger host halos, with each other, and with condensed baryonic objects all play a pivotal role in the survival of miniclusters~\cite{2017JETP..125..434D}.
In this paper, we quantify the degree to which interactions between miniclusters and stars can change the characteristics of AMCs today. We focus on the MW, where stars are abundant and constitute the dominant disruption mechanism~\cite{Berezinsky:2013fxa, Tinyakov:2015cgg}. In particular, we present a formalism that describes the reaction of an AMC's internal state to an interaction with a star. We then use this formalism to run Monte Carlo simulations of AMCs as they orbit the MW and interact with the stellar population. For computational simplicity, we make a number of simplifying assumptions. Firstly, we do not concurrently evolve AMCs through structure formation and stellar disruption. Secondly, we assume that the MW has been in a steady state since its formation. Despite these assumptions our results represent a fundamental step towards quantifying the importance of tidal stellar interactions for the distribution of AMCs in the MW. As we argue in \S~\ref{sec:caveats}, relaxing our assumptions will not change our overall conclusions but future work should quantitatively address these issues.

In addition to AMCs, axion stars (ASs) --- another class of axionic astrophysical object --- are expected to form and remain stable over cosmological times (in particular the dilute branch of axion stars~\cite{Visinelli:2017ooc}). Importantly, they may readily form within miniclusters, producing a central \textit{core}~\cite{Levkov:2018kau, Eggemeier:2019jsu, Chen:2020cef}. For simplicity, we do not simultaneously consider both AMCs and ASs. Instead, we make a cut on the minicluster parameter space in order to focus on those AMCs for which the density profile is known most reliably (as described in \S~\ref{sec:ASs}).

This paper is structured as follows: in Sec.~\ref{sec:AMCdistributions} we describe the initial distributions of AMCs in the MW that represent the starting point of our work. Section~\ref{sec:stripping} discusses the dynamics of successive stellar encounters. In Sec.~\ref{sec:Simulations} we discuss our Monte Carlo simulations and how to interpret the results. We then discuss the minicluster population today in Sec.~\ref{sec:AMCtoday} and how this applies to observational results in Sec.~\ref{sec:apps}. Finally, we discuss future work and conclude in Sec.~\ref{sec:conclusion}. Throughout this paper, we limit our discussion to axions which constitute 100\% of the DM and only consider the QCD axion. Specifically, we assume a KSVZ-like axion\footnote{KSVZ stands for Kim-Shifman-Vainshtein-Zakharov~\cite{Kim:1979if, Shifman:1979if}.} of mass $m_a = 20\,\mu$eV,\footnote{We choose the DM axion mass to agree with recent numerical simulations that also work under the assumption of a KSVZ axion~\cite{Klaer:2017qhr, Buschmann:2019icd}.} and we comment on results for different masses in Sec.~\ref{sec:conclusion}. Our results can be extended to other axion models, provided that the distributions of AMC masses and overdensities are modified accordingly. All code associated with this work (and the companion paper, Ref.~\cite{Edwards:2020afl}) is available online at \href{https://github.com/bradkav/axion-miniclusters/}{github.com/bradkav/axion-miniclusters} \cite{AMC_code}.

\section{Miniclusters in the Milky Way}
\label{sec:AMCdistributions}

Cold axions are produced in the early Universe through non-thermal processes at the time $t_{\rm osc}$ at which the axion field begins oscillating, when the Hubble rate is of the order of the axion mass.
The present number density of axions is given by
\begin{equation}
    \label{eq:axionnumberdensitytoday}
    n_a = \frac{\rho_{\rm osc}}{m_a}\,a_{\rm osc}^3\,,
\end{equation}
where $a$ is the scale factor (set to unity today),
and $\rho$ is the axion energy density. The subscript `osc' indicates the value of the parameter at $t_{\rm osc}$.
Equating the energy density of axions today to the observed DM abundance fixes the value of the axion mass. The computation of $\rho_{\rm osc}$ requires one to simulate the dynamics of the topological defects associated with the spontaneous breaking of the PQ symmetry, which is a significant challenge. Here, we follow recent literature on the subject and we fix $m_a = 20{\rm \,\mu eV}$~\cite{Klaer:2017qhr, Buschmann:2019icd}, although a wide range of masses is still possible~\cite{Gorghetto:2020qws}. 

An individual AMC can be characterized by an initial overdensity parameter $\delta$, discussed in \S~\ref{sec:overdensities}, and an initial mass described in \S~\ref{sec:HMF}. Due to the randomness of the initial conditions of the axion field over causally-connected patches, AMCs are formed with a range of overdensity parameters and masses. We note that predictions for the AMC properties are still under debate in the literature and we attempt to highlight these uncertainties throughout. We emphasize, however, that our framework can be straightforwardly re-cast under different assumptions for the initial distribution of AMCs, as we discuss in Sec.~\ref{sec:AMCtoday}.

\subsection{Distribution of Overdensities}
\label{sec:overdensities}

After the DM axion field has started to oscillate, its mean background density scales with temperature as $\bar \rho_a(T) = 3T_{\rm eq}\,s(T)/4$, where $s(T)$ is the entropy density at temperature $T$ and $T_{\rm eq}$ is the temperature at matter-radiation equality. A density fluctuation $\rho_a > \bar\rho_a$ decouples from the cosmological expansion at the temperature $T_\delta = (1+\delta)\,T_{\rm eq}$, where $\delta \equiv (\rho_a - \bar\rho_a)/\bar \rho_a$ is the overdensity parameter. After this, the overdense region undergoes non-linear gravitational collapse, becoming gravitationally bound into an AMC~\cite{Kolb:1993hw, Kolb:1995bu}. Assuming spherical collapse, the density of a virialized minicluster is~\cite{Kolb:1994fi}
\begin{equation}
	\rC(\delta) = 140\,\(1 + \delta\)\,\delta^3\,\rho_{\rm eq}\,,
	\label{eq:density_clusters}
\end{equation}
where $\rho_{\rm eq}$ is the average matter density at matter-radiation equality. The AMC density $\rC(\delta)$ does not depend on the interaction of the axion with matter, nor on the axion mass. For this reason, we expect that our results will be unchanged for AMCs formed from an axion-like field, as long as the axion-like field makes up the entirety of the DM.

The distribution of overdensities $\mathrm{d} f_{\rm AMC}/\mathrm{d}\delta$ can be assessed through numerical simulations~\cite{Kolb:1994fi, Buschmann:2019icd}. In this paper, we adopt the expression for $\mathrm{d} f_{\rm AMC}/\mathrm{d}\delta$ used in Ref.~\cite{Buschmann:2019icd} and we span the range of values $\delta \in \[0.1, 20\]$ (corresponding to characteristic densities $\rho_\mathrm{AMC} \in \[10^{2}, 2 \times 10^{10}\] \, M_\odot\,\mathrm{pc}^{-3}$). For completeness, we report the formula used in Appendix~\ref{sec:overdensity}. Mapping out the correlation between $\delta$ and the AMC mass $M_\mathrm{AMC}$ is currently challenging, as the simulations used to derive $\mathrm{d} f_{\rm AMC}/\mathrm{d}\delta$ stop at matter-radiation equality~\cite{Buschmann:2019icd}, while we are interested in the mass function in the late Universe. In the following, we assume that there is no correlation between $\mathrm{d}f_\mathrm{AMC}/\mathrm{d}\delta$ and the AMC mass distribution introduced below, though our formalism can be straightforwardly extended to incorporate such correlations.

\subsection{Initial Halo Mass Function}
\label{sec:HMF}

The characteristic comoving number density of AMCs
per logarithmic mass interval is described by the halo mass function (HMF). The HMF at matter-radiation equality (at redshift $z_{\rm eq}$) can be assessed by evolving the PQ field from the moment at which the PQ symmetry breaks until $z_{\rm eq}$~\cite{Vaquero:2018tib, Buschmann:2019icd}. The high-end tail of the HMF shows an exponential cutoff which, at recombination, is placed at around the largest mass of the AMCs, $M_{\rm max}(z_{\rm eq}) \approx M_0$~\cite{Vaquero:2018tib, Eggemeier:2019khm, Xiao:2021nkb}. The characteristic mass $M_0$ is associated with the axion energy density contained within a Hubble horizon at $t_{\rm osc}$~\cite{Visinelli:2018wza} 
\begin{equation}
    \label{eq:mcmass}
	M_0 = \frac{4\pi}{3}\(1+\delta\)\frac{\rho_{\rm osc}}{H_{\rm osc}^3} 
	\approx 10^{-11}M_\odot \(1+\delta\)\!\(\! \frac{20{\rm \,\mu eV}}{m_a}\!\)^{1/2}\,,
\end{equation}
where $M_\odot$ is the Solar mass.

Perturbations in the axion density continue to grow after matter-radiation equality, so that the HMF evolves under hierarchical structure formation. $N$-body simulations following AMC structures from recombination to $z\approx 99$ lead to an HMF $\mathrm{d}P/\mathrm{d}\ln M_\mathrm{AMC} \propto M_\mathrm{AMC}{}^\gamma$, with a characteristic slope $\gamma \sim -0.7$~\cite{Eggemeier:2019khm}.
This result corroborates the semi-analytic solution obtained by using the Press-Schechter formalism~\cite{Press:1973iz}, which finds that the mass function at late times scales as $M^{-0.68}$ for small masses, and as $M^{-0.35}$ for large masses, over the mass interval $10^{-15}\lesssim M/M_{\odot} \lesssim 10^{-9}$~\cite{Ellis:2020gtq}. Earlier work found the slope $\sim -0.5$~\cite{Fairbairn:2017sil, Fairbairn:2017dmf}. 

As structure formation proceeds, the high-end cutoff of the HMF evolves according to the Press-Schechter analysis, since AMCs of mass $\MC > M_0$ form through hierarchical structure formation from the early seeds of mass $\MC \leq M_0$. On the other hand, the assessment of the low-end mass cutoff is challenging~\cite[Sec.~5.2]{Niemeyer:2019aqm}.
Both the semi-analytic formalism and the numerical assessment of the low-end tail of the mass distribution $\MC \ll M_0$ show limitations due to a number of factors. For example, fluctuations in this regime are not Gaussian so the Press-Schechter formalism cannot be used; in addition, resolving the power spectrum at such small scales is a numerical challenge~\cite{Wang:2007he}. Note that neither the numerical simulations of the early Universe, nor the Press-Schechter formalism account for the possible presence of ASs, which we discuss in \S~\ref{sec:ASs}.

Here, we model the HMF at the present time $z = 0$ as
\begin{equation}
	\frac{{\rm d}P}{{\rm d}\ln \MC} = \frac{ \gamma}{M_{\rm max}{}^\gamma - M_{\rm min}{}^\gamma}\,\MC{}^\gamma\,,
	\label{eq:mcdistribution}
\end{equation}
where the expression is valid within the mass range $M_{\rm min} \leq \MC \leq M_{\rm max}$; otherwise we set the HMF to zero. We fix the characteristic slope to 
$\gamma = -0.7$ as suggested by recent simulations of the collapse and mergers of AMCs~\cite{Eggemeier:2019khm}. We adopt the HMF low-end cutoff $M_{\rm min}$ and high-end cutoff $M_{\rm max}$ from Refs.~\cite{Fairbairn:2017sil, Fairbairn:2017dmf}.\footnote{We note that recently Ref.~\cite{Xiao:2021nkb} found a HMF with an overall shift to lower masses. As we will show below, the disruption process is approximately independent of the AMC mass; this shift will therefore not have a substantial effect on our results.} The low-end tail of the HMF distribution is cut off at the mass $M_{\rm min}$, which at the time of AMC formation is proportional to the axion Jeans mass. These smallest collapsed miniclusters then grow slowly to $z=0$ today.

The high-end cutoff of the HMF arises from the fact that the largest overdensities in the initial Gaussian density field are exponentially suppressed. The exact value of $M_{\rm max}$ is not important in determining the properties of the AMC distribution, since the HMF is peaked towards low values of the mass with negligible contributions from masses $\MC \gg M_0$. 
\begin{align}
\label{eq:Mminmax}
    \begin{split}
        M_\mathrm{min} &= 3.3 \times 10^{-19} \,M_\odot \,,\\
        M_\mathrm{max} &= 5.1 \times 10^{-5} \,M_\odot\,.\\
    \end{split}
\end{align}
Note that the lower end of this mass range will be suppressed by our AS cut, as described in \S~\ref{sec:ASs}. The characteristic radius $\RC$ for an AMC of mass $\MC$ is of the order of
\begin{align}
    \begin{split}
    \RC &\sim \left(\frac{3\MC}{4\pi \rC(\delta)}\right)^{1/3} \\
    &\approx 1.4\times 10^{11}{\rm \, m}\,\(\frac{\MC}{10^{-10}\,M_\odot}\)^{1/3}\,.
    \label{eq:MCradius}
    \end{split}
\end{align}

The HMF in Eq.~\eqref{eq:mcdistribution} has been obtained without considering the effects of the tidal stripping of AMCs due to nearby stars in the MW or due to the mean Galactic field. The distribution in Eq.~\eqref{eq:mcdistribution} gives us the {\it initial} HMF. In Sec.~\ref{sec:stripping}, we assess the effects of tidal stripping on the population of AMCs, which effectively modifies the initial HMF to yield the \textit{true} HMF today.
Because the HMF is a falling Power-law, the lightest AMCs will dominate the MW population, meaning that our results could in principle be sensitive to the low-mass cut-off $M_\mathrm{min}$.
While $M_\mathrm{min}$ lies below the minimum mass that passes our AS cut criteria, it may still affect the AMC population through its influence on the normalization of the HMF.

\subsection{Distribution of AMCs in the Galaxy}
\label{sec:Distribution_AMCs}

Given the DM density profile in the MW $\rho_{\rm DM}(r)$, we model the spatial distribution of the number density of AMCs as
\begin{equation}
    \label{eq:MCdistribution}
    \nC(r) = \fC \frac{\rho_{\rm DM}(r)}{\langle \MC\rangle}\,,
\end{equation}
where $\langle \MC \rangle$ is the mean AMC mass {\it before} disruption is accounted for. Using the HMF in Eq.~\eqref{eq:mcdistribution}, we obtain the value $\langle \MC \rangle = 1.4\times 10^{-14}\,M_\odot$. 
Here, we set $\fC = 100\%$ (though all our results can be trivially rescaled).

The distribution of the DM density in the Galaxy $\rho_{\rm DM}(r)$ is modelled according to a Navarro-Frenk-White (NFW) density profile~\cite{Navarro:1996gj},
\begin{equation}
    \rho_{\rm NFW}(r) = \frac{\rho_s}{(r/r_s)(1 + r/r_s)^2}\,,
    \label{eq:density_NFW}
\end{equation}
where we set the parameters $\rho_s = 0.014\,M_\odot{\rm \,pc}^{-3}$ and $r_s = 16.1\,$kpc~\cite{Nesti:2013uwa}.

\subsection{Density Profiles of AMCs}
\label{sec:PhaseSpace}

\begin{figure}[t!]
\includegraphics[width=0.49\textwidth]{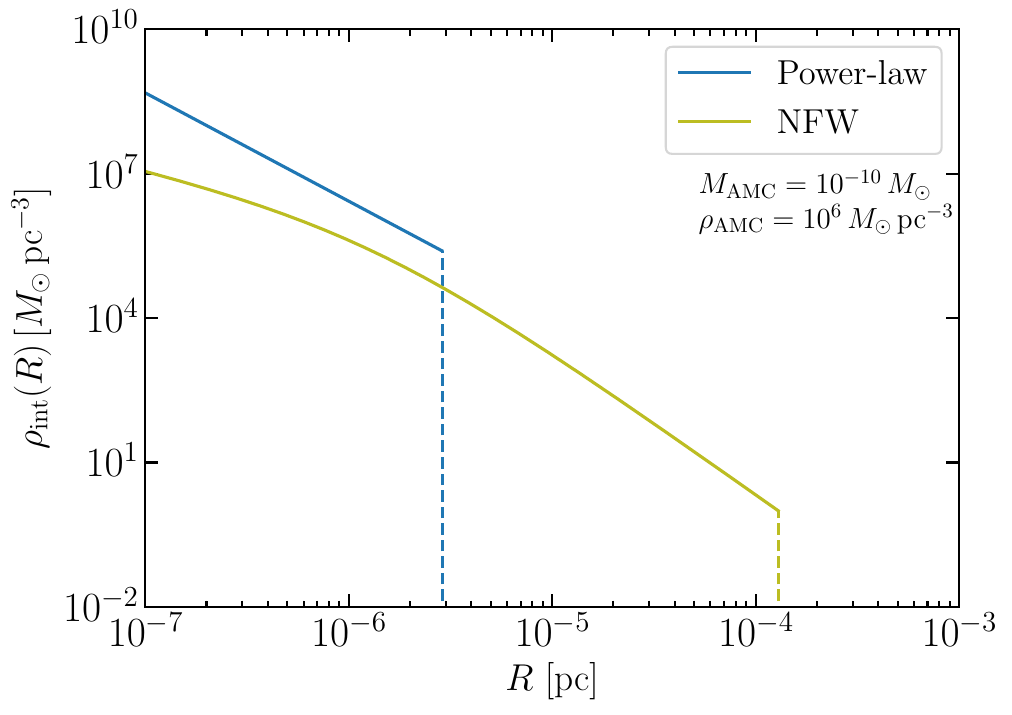}
	\caption{Models for the internal density profile of AMCs which we consider in this work: Power-law, Eq.~\eqref{eq:PLprofile}, and NFW, Eq.~\eqref{eq:density_NFW}. Vertical dashed lines show the truncation radii $R_\mathrm{AMC}$. We fix the characteristic mass and density to $M_{\mathrm{AMC}}=10^{-10} \, M_{\odot}$ and $\rho_{\mathrm{AMC}}=10^6\, M_{\odot} \,\mathrm{pc}^{-3}$ ($\delta \approx 1.55$) respectively. In the NFW case, the mean density is much lower and the AMC is much larger.}
	\label{fig:AMC_densityprofiles}
\end{figure}

For the internal density profile of the AMCs $\rho_{\rm int}(R)$, we consider two different models, namely {\bf i)} a self-similar Power-law (PL) profile, and {\bf ii)} an NFW profile as in Eq.~\eqref{eq:density_NFW}. These density profiles are illustrated in Fig.~\ref{fig:AMC_densityprofiles}.\footnote{In the rest of the paper, we heroically endeavour to use lower-case $r$ for galactocentric radii and upper-case $R$ for AMC radii.} 
The density profile of an AMC for case {\bf i)} is described by~\cite{OHare:2017yze,Fairbairn:2017sil}
\begin{equation}
    \label{eq:PLprofile}
    \rho_\mathrm{int}^{\mathrm{PL}}(R) = \rho_s \left(\frac{r_s}{R}\right)^{9/4}\,{\rm \Theta}\left( R_\mathrm{AMC}^{\rm PL} - R \right)\,,
\end{equation}
 where ${\rm \Theta}\left(x\right)$ is the Heaviside step function. We truncate the PL profile at a radius 
\begin{equation}
    \label{eq:RmaxPL}
    R_\mathrm{AMC}^{\rm PL} = \left(\frac{3 \MC}{4 \pi \rC(\delta)}\right)^{1/3}\,.
\end{equation}
We fix $\rho_s r_s^{9/4} = \rC(\delta) (R_\mathrm{AMC}^{\rm PL})^{9/4}/4$~\cite{Fairbairn:2017sil}, to give mean density $\rho_\mathrm{AMC}(\delta)$ and the correct total mass for the AMC. 
Such PL profiles are expected from models of secondary infall~\cite{Bertschinger:1985pd} and have also been observed in numerical simulations of the gravitational collapse of AMCs~\cite{Zurek:2006sy}. The PL profile likely describes the density of AMCs at formation and therefore would be most suitable for the lightest AMCs which have not undergone growth through mergers.  

The most massive AMCs are formed hierarchically through mergers of smaller AMCs, as is generically expected for cold DM substructures~\cite{Fairbairn:2017dmf,Fairbairn:2017sil,Eggemeier:2019khm}. This motivates model {\bf ii)}, in which the density is described by an NFW profile, defined in Eq.~\eqref{eq:density_NFW}. The exact correspondence between ($M_\mathrm{AMC}$, $\delta$) and the NFW parameters is somewhat arbitrary. We follow Ref.~\cite{Fairbairn:2017sil} and make the identifications $\rho_s = \rC(\delta)$ and
\begin{equation}
\label{eq:NFWscaleradius}
	r_s = \(\frac{\MC}{4 \pi \rC(\delta) f_{\rm NFW}(c)}\)^{1/3}\,,
\end{equation}
where the function $f_{\rm NFW}(c) = \ln(1 + c) - c/(1+c)$ is defined in terms of a `truncation' parameter $c = R_\mathrm{AMC}/r_s$, which we fix to $c = 100$.\footnote{This truncation parameter is analogous to the concentration parameter used to characterize isolated NFW halos, and which relates the virial radius of the halo to its scale radius, $c = R_\mathrm{vir}/r_s$. We will sometimes therefore refer to $c$ as the `concentration' of the AMC.} The AMC profile is truncated at the radius
\begin{equation}
    \label{eq:RmaxNFW}
	R_\mathrm{AMC}^{\rm NFW} = c\, r_s \sim  6.5\times 10^{12}{\rm\, m}\(\frac{\MC}{10^{-10}M_\odot}\)^{1/3}\,,
\end{equation}    
where the numerical result holds for $\delta = 1$. With these choices, the mean density enclosed within $r_s$ is $3 f_\mathrm{NFW}(1) \rho_\mathrm{AMC} (\delta)\approx 0.58 \rho_\mathrm{AMC}(\delta)$ and the total mass enclosed within $R_\mathrm{AMC}$ is $M_\mathrm{AMC}$.

Numerical simulations suggest AMC concentrations of $c \sim \mathcal{O}(100)$ at $z = 99$~\cite{Eggemeier:2019khm}, growing roughly as $(1+z)^{-1}$ to $c \sim \mathcal{O}(10^4)$ today~\cite{Ellis:2020gtq}. However, AMCs with such large concentrations would have a density in their outskirts which is many orders of magnitude lower than that of the host halo of the MW. We therefore fix $c = 100$ to account for the fact that such diffuse AMCs would have been rapidly tidally truncated. This tidal truncation will lead to a reduction in the AMC mass compared to the \textit{initial} mass described in \S~\ref{sec:HMF}. As described in more detail in \S~\ref{sec:caveats} and Appendix~\ref{app:host_halo}, this amounts to a mass loss of 5-40\% (depending on the initial AMC mass). For NFW AMCs, we therefore correct the initial HMF to account for this mass loss, as described in \S~\ref{sec:PhysicalProperties}. 
Note also that as we show in Appendix~\ref{app:NFWassumptions}, fixing the truncation parameter to $c = 10^4$ instead should have a minimal impact on our formalism and results.

Even fixing $c = 100$, the above choices lead to NFW miniclusters which are much more dilute than the PL case, allowing us to also explore a more conservative scenario. For a given $\MC$ and $\rho_\mathrm{AMC}$, AMCs described by this NFW profile (with $c = 100$) will have a \textit{mean} internal density which is $\mathcal{O}(10^5)$ times lower than the corresponding PL profile, as illustrated in Fig.~\ref{fig:AMC_densityprofiles}.

Recent N-body simulation~\cite{Eggemeier:2019khm} suggest that the transition from direct collapse (PL-like profiles) to hierarchical structure formation (NFW-like profiles) should occur at around $M\sim 10^{-13}\,M_\odot$ for an axion of mass $m_a=20{\rm \,\mu eV}$. 
However, dedicated cosmological simulations are required to confirm the detailed behaviour of AMC density profiles as a function of $\MC$. We therefore perform our analysis assuming that either all AMCs have PL profiles or that all have NFW profiles, in the spirit of bracketing the uncertainties on the final AMC properties.

 \subsection{Axion Stars}
\label{sec:ASs}

Non-relativisitic ASs are described by solitonic solutions to the Schr\"{o}dinger-Poisson equation and are expected to form in the central regions of AMCs in the right conditions. In particular, the central density of an AMC must be high enough to allow two-to-two processes to cool their inner core and lead to the formation of a Bose-Einstein condensate~\cite{Kolb:1993zz, Seidel:1993zk}. This process has been observed in recent numerical simulations~\cite{Levkov:2018kau, Eggemeier:2019jsu, Chen:2020cef} and has shown a characteristic core-halo mass relation~\cite{Schive:2014hza}
\begin{equation}
M_{\rm AS} = 1.56 \times 10^{-13}\, M_\odot \left(\frac{20{\rm \,\mu eV}}{m_a}\right)\left(\frac{ \MC }{1 \,M_\odot}\right)^{1 / 3}\,,
\end{equation}
where we have evaluated the expression today and ignored $\mathcal{O}(1)$ factors. The corresponding radius is given by~\cite{Schive:2014hza}
\begin{align}
    \begin{split}
        R_{\rm AS} &= 3.85\times 10^{-8}{\rm \,m} \(\frac{20{\rm \,\mu eV}}{m_a}\)^2\,\left(\frac{M_\odot}{M_{\rm AS}}\right)\\
        &= 2.47 \times 10^5\,\mathrm{m} \left(\frac{20{\rm \,\mu eV}}{m_a}\right)\left(\frac{ \MC }{1 \,M_\odot}\right)^{-1 / 3}\,.
    \end{split}
    \label{eq:axionstarradius}
\end{align}
The inverse relationship between the AS's mass and radius leads to a problematic scenario for low-mass AMCs in which the central AS's radius would be \textit{larger} than that of the corresponding AMCs. To avoid this unphysical description of an AMC, we perform a cut on the overall population in which we only consider miniclusters with a radius larger than the radius of the corresponding AS at its center, i.e.
\begin{equation}
R_\mathrm{AMC}(\delta) > R_{\rm AS}\,.
\end{equation}
These results will be referred to as the `AS cut'.

For the smallest overdensity parameter that we consider $\delta = 0.1$, we find that no AMCs with masses below $5.0 \times 10^{-16} \,M_\odot$ pass the AS cut for PL profiles, while no AMCs below $1.6 \times 10^{-18} \,M_\odot$ pass in the case of NFW profiles. In both cases, these minimum masses exceed the value of $M_\mathrm{min}$ in Eq.~\eqref{eq:Mminmax}. Starting from the \textit{initial} population of AMCs described in this section, we then find the fraction of AMCs which pass the AS cut is 
$f_\mathrm{cut}^\mathrm{PL} = 2.7 \times 10^{-4}$ for PL density profiles and $f_\mathrm{cut}^\mathrm{NFW} = 1.5 \times 10^{-2}$ for NFW profiles. The difference between the two density profiles arises because for a fixed mass $M_\mathrm{AMC}$ and characteristic density $\rho_\mathrm{AMC}$, PL profiles are more compact and the AS radius in Eq.~\eqref{eq:axionstarradius} is more likely to exceed the AMC radius.

We emphasize that current numerical simulations (for example, Refs.~\cite{Buschmann:2019icd,Eggemeier:2019jsu,Eggemeier:2019khm}) do not have sufficient resolution to observe the formation of ASs in the lightest AMCs and therefore their existence and evolution has not yet been confirmed. However, our aim is to cut out AMC-AS systems which are likely to be most problematic. Even with this cut, it is also possible that the central density core produced by the presence of an AS may affect the stability of AMCs to tidal perturbations. The treatment of light AMC-AS systems requires dedicated study and is left to future work.

\section{Tidal Stripping of Axion Miniclusters}
\label{sec:stripping}

AMCs can be disrupted by their encounters with stars~\cite{Zhao:2005py} as well as by tidal interactions with the gravitational field of the disk~\cite{Berezinsky:2013fxa}. In this section,  we aim to model the interactions of stars with AMCs. Importantly, we model and track all interactions, including those that do \textit{not} lead to the total disruption of an AMC. Through many successive weak interactions, these AMC's can lose mass and potentially have greatly enlarged radii. This population of {\it perturbed} AMCs may result in quantitatively distinct observational signatures when compared to an unperturbed population (see Sec.~\ref{sec:apps} and Ref.~\cite{Edwards:2020afl}).\footnote{These encounters may leave a stream of axions behind them which can also lead to features in direct detection experiments~\cite{Tinyakov:2015cgg, Knirck:2018knd}, but here we focus on the properties of surviving AMCs.}

First, we describe how to treat an individual AMC going through a series of interactions. We then discuss in Sec.~\ref{sec:Simulations} our Monte Carlo procedure to model a population of AMCs being perturbed. 

\subsection{Encounter Dynamics}\label{sec:interactions}

\begin{figure}[t!]
\includegraphics[width=0.45\textwidth]{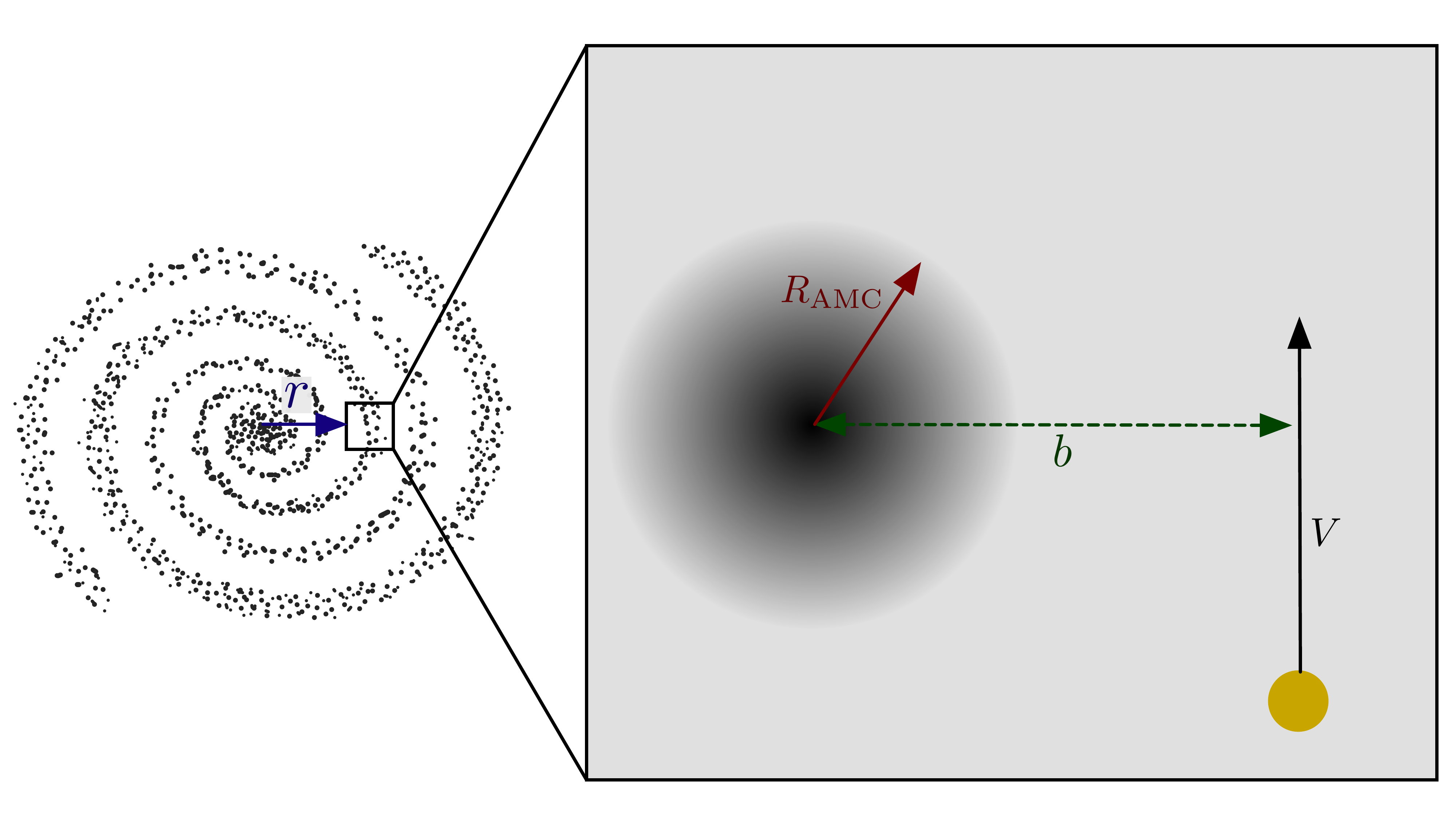}
	\caption{Illustration of an AMC-star encounter taking place at a distance $r$ from the Galactic center. An AMC with radius $R_\mathrm{AMC}$ is passed by a star of mass $M_\star$ with impact parameter $b$ and relative velocity $V$. The energy $\Delta E$ injected into the AMC is given in Eq.~\eqref{eq:energy_perturbation}.
	}
	\label{fig:interaction_cartoon}
\end{figure}

Stars are dense objects with relatively small radii. Similarly, AMCs are small, meaning that the large majority of encounters will occur when the separation between these objects is significantly larger than their physical size. We therefore work in the `distant-tide' approximation~\cite{BinneyTremain:2008}. In this approximation, a minicluster of mass $\MC$ going through an encounter with a stellar object would increase its internal energy by a quantity~\cite{1958ApJ...127...17S} (see also Refs.~\cite{Green:2006hh, Schneider:2010jr,Hertzberg:2019exb,Delos:2019tsl}):
\begin{equation}
	\Delta E \approx \(\frac{2GM_{\star}}{b^2 V}\)^2 \frac{\MC\,\langle R^2 \rangle}{3}\,,
	\label{eq:energy_perturbation}
\end{equation}
where $M_{\star}$ is the mass of the stellar object, $b$ is the impact parameter of the interaction, $V$ is the relative velocity between the objects, and the mean squared radius $\langle R^2 \rangle$ accounts for the mass distribution inside the AMCs~\cite{Green:2006hh}. We parametrize the mean squared radius as $\langle R^2 \rangle = \alpha^2 R_\mathrm{AMC}{}^2$, with $\alpha^2 = 3/11 \approx 0.27$ for the PL profile and $\alpha^2 \approx 0.13$ for the NFW profile. Such an encounter is illustrated in Fig.~\ref{fig:interaction_cartoon}.

The size of the energy injection, as described by Eq.~\eqref{eq:energy_perturbation}, should be compared with the binding energy of the AMC, which we write as $E_\mathrm{bind} = \beta G M_\mathrm{AMC}{}^2/R_\mathrm{AMC}$. The $\mathcal{O}(1)$ prefactor $\beta$ depends on the internal density profile for which we find $\beta = 1.5$ for the PL profile and $\beta = 3.46$ for the NFW profile. There are then two distinct regimes for the energy injection:\footnote{We only split these two regimes for the purposes of discussion. Computationally, both are treated in the same way.} 
\begin{itemize}
     \item An encounter with a sufficiently small impact parameter will inject more energy than the binding energy $E_\mathrm{bind}$ of the AMC leading to complete disruption.
     \item An encounter with a large impact parameter that simply injects energy into the AMC but does not completely unbind it.
\end{itemize}
   The first regime ($\Delta E \gtrsim E_\mathrm{bind}$) can be re-expressed as $b \lesssim b_{\rm min}$, where we have defined the minimal impact parameter that does not entirely disrupt the minicluster as
\begin{equation}
	b_{\rm min}(\delta) \approx \sqrt{\frac{M_{\star}}{V}}\,\(\frac{\alpha^2 \,G}{\beta \,\pi\, \bar{\rho }(\delta)}\)^{1/4}\,.
    \label{eq:impactparameter}
\end{equation}
Here, $\bar{\rho} = 3 M_\mathrm{AMC}/(4 \pi R_\mathrm{AMC}(\delta)^3)$ is the mean density of the AMC. An encounter between a PL AMC with $\delta = 1$ and a perturbing object of mass $M_{\star} = 1\,M_{\odot}$ with a relative velocity $V = 10^{-3}\,c$ gives $b_{\rm min} \approx 0.01\,$pc, which is much larger than the typical size of an AMC, as is required by the distant-tide approximation. 
Note that this expression depends only on the density of the minicluster, and not on its size or mass separately. Indeed, the fractional energy injected $\Delta E/E_\mathrm{bind}$ depends on the AMC properties only through the mean density, and we therefore expect that the behaviour of the AMCs under perturbations should be independent of $M_\mathrm{AMC}$. As pointed out in \S~\ref{sec:PhaseSpace}, for a given mass $M_\mathrm{AMC}$ and overdensity $\delta$, the mean density of an AMC is significantly lower for NFW profiles than for PL profiles. As we will see, AMCs with NFW profiles are much more easily disrupted than their PL counterparts. 

The second regime occurs for larger values of the impact parameter $b > b_{\rm min}$. In this regime, a single encounter does not completely unbind the AMC, but energy injected through multiple encounters can lead to mass-loss or a change in radius and may eventually disrupt the AMC. We study this second regime in more detail below.

\subsection{Perturbing the Miniclusters}
\label{sec:Perturbation}

To estimate the mass loss from a minicluster when it is perturbed, we study the evolution of the phase space distribution function of axions in the minicluster:
\begin{equation}
    f(\mathcal{E}) \equiv m_a  \,\frac{\mathrm{d}N}{\mathrm{d}^3\mathbf{R}\,\mathrm{d}^3\mathbf{v}}\,.
\end{equation}
For isotropic, spherically symmetric distributions of particles, the distribution function depends only on their specific relative energy 
\begin{equation}
    \mathcal{E} = -\frac{E}{m_a} = \Psi(R) - v^2/2\,,
\end{equation}
where $\Psi(R) = - \Phi(R)$ is the gravitational potential relative to the boundary at infinity~\cite[Ch.~4.3]{BinneyTremain:2008}.

For spherically symmetric systems in equilibrium, the potential is a monotonic function of the radius $R$, meaning that the density profile can be expressed as a function of $\Psi$, $\rho(R) = \rho(\Psi(R))$. The distribution function can then be determined from the density profile using the Eddington inversion method~\cite[p. 290]{BinneyTremain:2008}:
\begin{equation}
    f(\mathcal{E}) \equiv \frac{1}{\sqrt{8}\pi^2} \int_{0}^{\mathcal{E}} \frac{1}{\sqrt{\mathcal{E} - \Psi}}\frac{\mathrm{d}^2\rho}{\mathrm{d}\Psi^2}\,\mathrm{d}\Psi\,.
\end{equation}
As discussed in \S~\ref{sec:PhaseSpace}, we will consider two possible density profiles for the AMCs: Power-law (PL) and NFW. For the PL  profile, the distribution function can be computed analytically (see e.g.~Ref.~\cite{Gondolo:1999ef}) while the NFW distribution must be computed numerically. See Appendix~\ref{app:distributionfunctions} for more details.

Consider then a perturbation to the minicluster of total size $\Delta E$. Given that the critical impact parameter for disrupting the minicluster is already much larger than the minicluster size, we will assume that $b \gg R_\mathrm{AMC}$. Under this condition, the average energy injected per unit mass increases with distance from the AMC center as $R^2$~\cite{1958ApJ...127...17S, Green:2006hh} and can be written:
\begin{equation}
    \label{eq:DeltaE_R}
    \Delta \mathcal{E}(R) = -\frac{\Delta E}{M_\mathrm{AMC}}\frac{R^2}{\langle R^2 \rangle}\,.
\end{equation}
Particles with $\mathcal{E} < 0$ immediately after the perturbation can be considered unbound; numerical simulations of the disruption of stellar clusters suggest that the subsequent relaxation of the system should not substantially change the fraction of particles which are unbound~\cite{Gieles:2006ch,2020arXiv200906643M}. So in order to compute the mass loss we need only calculate the minicluster mass in particles with energy $\mathcal{E} < \Delta \mathcal{E}$.
This is given by:
\begin{widetext}
\begin{align}
\label{eq:deltaM}
    \begin{split}
         \Delta M = M(< \Delta\mathcal{E}) &= \int_{\mathcal{E} < \Delta \mathcal{E}(R)}  \mathrm{d}^3\mathbf{R}\,\mathrm{d}^3\mathbf{v}\,f(\mathcal{E})\\
         &= 16 \pi^2 \int_0^{R_\mathrm{AMC}} R^2 \,\mathrm{d}R \int_0^{v_\mathrm{max}(R)} v^2\,\mathrm{d}v\, f(\mathcal{E}) \Theta(\Delta\mathcal{E}(R) - \mathcal{E})\\
         &= 16 \pi^2 \int_0^{R_\mathrm{AMC}} R^2 \,\mathrm{d}R \int_0^{\mathrm{min}[\Delta\mathcal{E}(R), \Psi(R)]} \,\mathrm{d}\mathcal{E}\,\sqrt{2 (\Psi(R) - \mathcal{E})} f(\mathcal{E})\,,
    \end{split}
\end{align}
\end{widetext}
where the escape speed is $v_\mathrm{max}(R) = \sqrt{2 \Psi(R)}$ and we have also used $v = \sqrt{2(\Psi(R) - \mathcal{E})}$ and $\mathrm{d}\mathcal{E} =  - v \,\mathrm{d}v$. All  calculations in this section are done assuming a fixed potential $\Psi(R)$ as given just before the interaction, which is justified in the limit where the mass loss is small and mostly happening at the outskirts of the object.

Note that we have defined the distribution function $f(\mathcal{E})$ for an isolated AMC which extends to infinity. In practice, in Eq.~\eqref{eq:deltaM}, we implement a hard truncation of the AMC at a radius $R = R_\mathrm{AMC}$. We assume therefore that particles at $R > R_\mathrm{AMC}$ are no longer bound to the AMC but are instead bound to the diffuse halo of the MW. Note also that our definition of the distribution function is not strictly consistent with a truncated density profile, as we have calculated the potential assuming that the AMC extends to infinity (for convenience). Physically, this means that particles near $R = R_\mathrm{max}$ are moving more quickly (and are therefore more easily unbound) than they would be in a self-consistent model for the miniclusters. We therefore consider this calculation as conservative from the perspective of AMC disruption. Nonetheless, the error induced by this approximation should be small because, as we will see, these density profiles are close to being in virial equilibrium.

\begin{figure}[t!]
\includegraphics[width=0.5\textwidth]{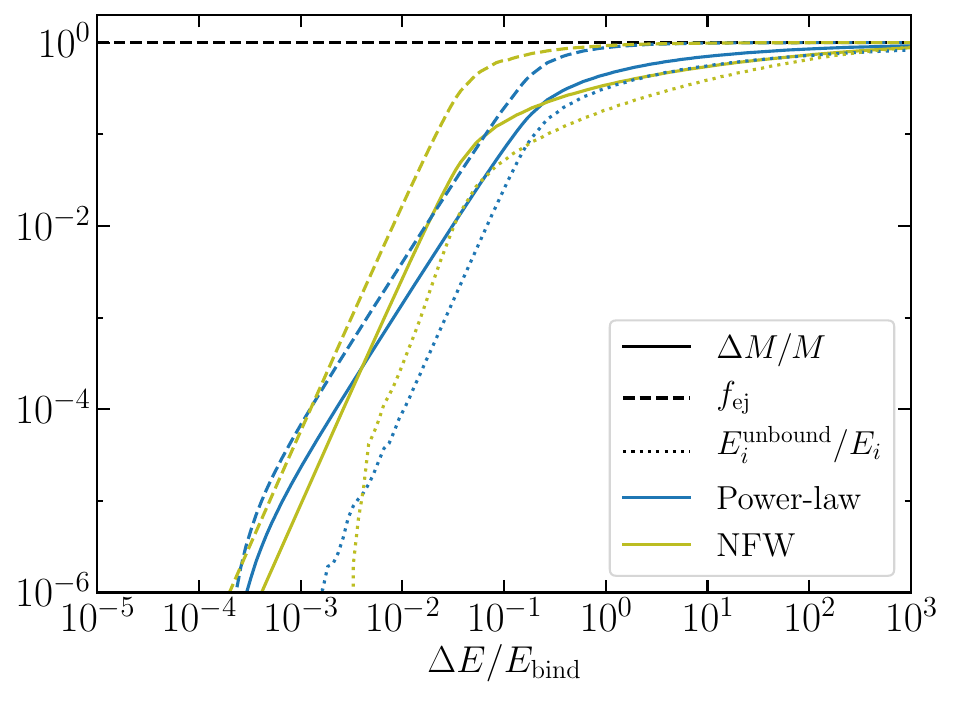}
	\caption{Fractional mass loss (solid lines) for AMCs with Power-law and NFW profiles as a function of the size of the perturbation $\Delta E$, in units of the binding energy $E_\mathrm{bind}$. Dashed lines show the fraction of the injected energy which is carried away by the ejected mass $f_\mathrm{ej}$, while dotted lines show the fraction of the initial AMC energy stored in particles which are eventually unbound in the interaction.} 
	\label{fig:MassLoss}
\end{figure}

In Fig.~\ref{fig:MassLoss}, we plot the mass loss as a function of the size of the perturbation $\Delta E$, expressed in terms of the binding energy $E_\mathrm{bind} = \beta G M_\mathrm{AMC}{}^2/R_\mathrm{AMC}$. 
The fraction of mass lost in an encounter grows with the size of the perturbation, tending slowly to $\Delta M \sim M$ for $\Delta E \gg E_\mathrm{bind}$. The `flattening' in $\Delta M /M$ occurs because energy is predominantly injected into particles in the outskirts of the AMC which are only loosely bound. Very large amounts of energy are required to strip away the tightly bound particles close to the center. 
Our calculations are in line with results from $N$-body simulations of stellar clusters, which show a mass loss of $20$-$30$\% for energy injections $\Delta E \sim E_\mathrm{bind}$ (see e.g.~Fig.~6 of Ref.~\cite{2020arXiv200906643M}).

Once some mass has been stripped away from the minicluster, we must compute the properties of the surviving object. The total energy of the minicluster is: 
\begin{equation}
\label{eq:AMCenergy1}
    E_\mathrm{total} = \frac{1}{2}\MC \,\sigma^2  - E_\mathrm{bind}\,.
\end{equation}
The velocity dispersion squared $\sigma^2$ can be computed from the distribution function and may be parametrized as $\sigma^2 = \kappa G \MC/\RC$. The total energy can then be written 
\begin{equation}
\label{eq:AMCenergy2}
    E_\mathrm{total} = \left( \frac{\kappa}{2} - \beta \right) \frac{G \MC^2}{\RC}  = \left(\frac{\kappa}{2\beta} - 1\right) E_\mathrm{bind}\,,
\end{equation}
allowing the energy of the AMC to be related to its mass and radius. Objects in virial equilibrium should have $\kappa = \beta$. As we have noted above, the artificially truncated profiles we consider are not strictly in equilibrium, leading to values of $\kappa = 1.15 \,\beta$ for PL profiles and $\kappa = 1.02 \,\beta$ for NFW profiles. In Table~\ref{tabParams} we collect the numerical values of the prefactors obtained for each expression.

\begin{table}[tb!]
\def\arraystretch{1.5}
	\begin{tabular}{cccc}
    \cline{2-4}
    \hline\hline
		 & PL & NFW & Expression \\ 
		 \hline
		$\alpha^2$ & 0.27 & 0.13 & $\langle R^2 \rangle = \alpha^2 R_\mathrm{AMC}{}^2$ \\ 
		$\beta$ & 1.5 & 3.47 & $E_\mathrm{bind} = \beta\, G M_\mathrm{AMC}{}^2/R_\mathrm{AMC}$ \\ 
		$\kappa$ & 1.73 & 3.54 & $\sigma^2 = \kappa\, G \MC/\RC$ \\ \hline\hline
	\end{tabular}
	\caption{Numerical values of the prefactors used for characterizing the AMC properties: the mean-squared radius $\langle R^2 \rangle$; binding energy $E_\mathrm{bind}$; and velocity dispersion squared $\sigma^2$.}
	\label{tabParams}
\end{table}

Energy conservation implies that
\begin{align}
 E_i^\mathrm{bound} + \Delta E &= E_f^\mathrm{bound} +  E_f^\mathrm{unbound}\,,
\end{align}
where the subscripts $i$ and $f$ denote quantities defined just before and after the interaction. Superscripts `bound' or `unbound' refer to the respective subsets of particles, as defined through Eq.~\eqref{eq:deltaM}, and we use $E_i^\mathrm{bound} \equiv E_i$ and $E_f^\mathrm{bound} \equiv E_f$ for clarity.   We assume that unbound particles are removed instantaneously to infinity. Energies include both the kinetic energy of the particles as well as their potential energy.

We can estimate $E_f^\mathrm{unbound}$ by taking the initial (pre-interaction) energy of the subset of particles that is going to be unbound $E_i^\mathrm{unbound}$ and adding the energy that is transferred to these particles during the interaction.
This yields
\begin{align}
    E_f^\mathrm{unbound} = E_i^\mathrm{unbound} + f_\mathrm{ej} \Delta E\,,
\end{align}
where $f_\mathrm{ej}$ is the fraction of the total injected energy that goes to unbound particles.
Putting these components together, we can write the final energy of the AMC after collision as
\begin{align}
\label{eq:E_f}
E_f = E_i + (1-f_\mathrm{ej})\Delta E  - E_i^\mathrm{unbound}\,.
\end{align}

The fraction $f_\mathrm{ej}$ can be calculated by performing a similar calculation as for the mass loss in Eq.~\eqref{eq:deltaM}, but weighting the integral by the amount of energy injected at each radius $\Delta \mathcal{E}(R)$, as given in Eq.~\eqref{eq:DeltaE_R}.  We find that $f_\mathrm{ej}$ depends on the size of the perturbation $\Delta E$ and is typically a factor of a few times larger than the fraction of mass ejected $\Delta M/M_\mathrm{AMC}$, as illustrated in Fig.~\ref{fig:MassLoss}. The initial energy of the particles which will eventually be unbound $E_i^\mathrm{unbound}$ can be taken as the sum of the kinetic energy of these particles plus the change in the binding energy from removing these particles. The final binding energy is calculated self-consistently from the density profile immediately after the interaction (see Appendix~\ref{app:distributionfunctions} for more details).

We assume that after the perturbation, the AMC will relax on a short timescale to have the same density profile (and for NFW profiles, the same truncation parameter $c = 100$), but described by a new mass and radius. This assumption is made for computational simplicity --- however, there is some indication from N-body simulations that perturbed DM substructures do retain a universal density profile~\cite{Delos:2019tsl, 2020arXiv200906643M}.\footnote{Note, however that Ref.~\cite{Delos:2019tsl} uses different definitions for binding energy and energy injections.} The final mass is $\MC^f = \MC^i - \Delta M$, while the final radius can be calculated from the total energy using Eq.~\eqref{eq:AMCenergy2}:
\begin{equation}
    \RC^f = \left(\frac{\kappa}{2} - \beta\right) \frac{G \left(\MC^f\right)^2}{E_f}\,.
\end{equation}
We note that the internal relaxation time scale $t_\mathrm{rel} \sim R_\mathrm{AMC}/\sigma_v \sim 10^4\,\mathrm{yr}$ of the AMCs is several orders of magnitude shorter than the average time between substantial encounters, $ \Delta t_\mathrm{enc} \sim t_\mathrm{MW}/N_\mathrm{enc} \sim \mathcal{O}(10^6)\,\mathrm{yr}$, where $t_\mathrm{MW} \approx 13.5 \times 10^{9}\,\mathrm{yr}$ is the age of the MW and $N_\mathrm{enc} \sim 10^4$ is the typical number of encounters (see Fig.~\ref{fig:Ninteractions}). We therefore assume that there should be sufficient time for AMCs to relax between successive encounters.

\begin{figure*}[t!]
\includegraphics[width=0.49\textwidth]{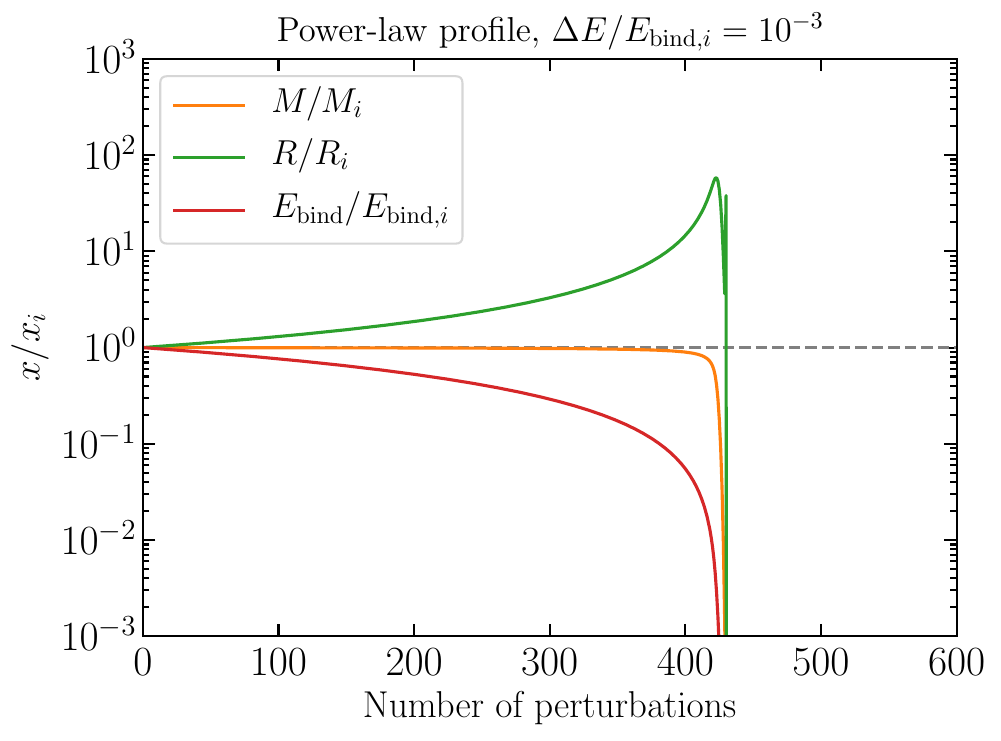}
\includegraphics[width=0.49\textwidth]{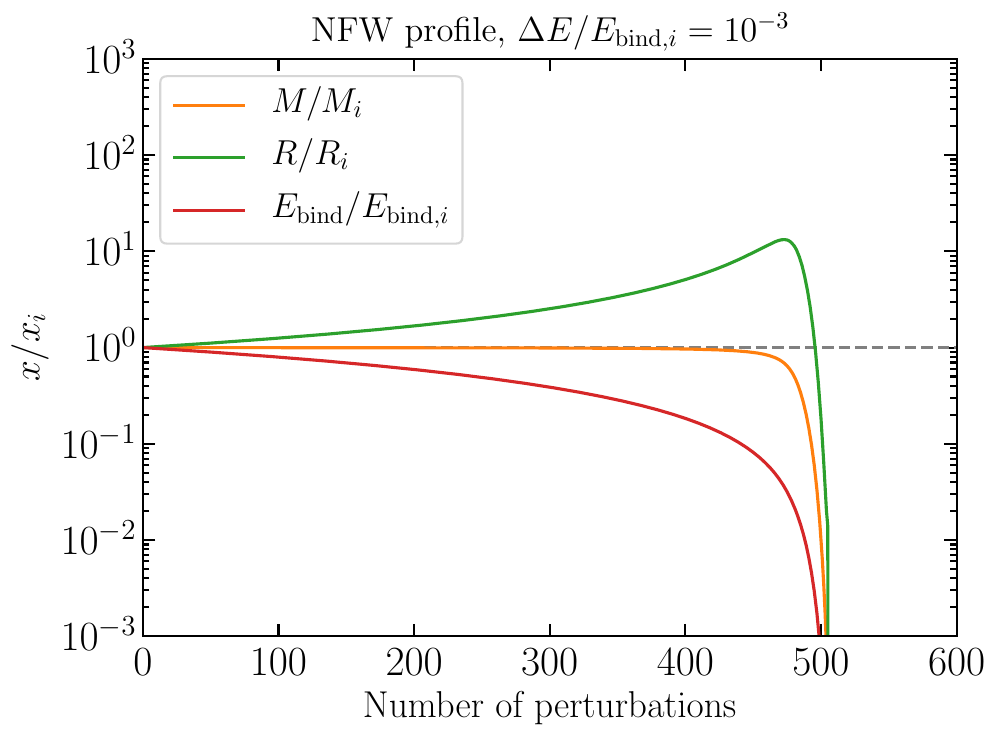}
	\caption{Fractional change in the minicluster properties $x = \{M, R, E_\mathrm{bind}\}$ under repeated perturbations for AMCs with Power-law profiles (left panel) and NFW profiles (right panel). We fix the size of each perturbation to be 1000 times smaller than the initial binding energy, $\Delta E = 10^{-3} E_{\mathrm{bind}, i}$.}
	\label{fig:Perturbations}
\end{figure*}

The assumption that the AMCs will have the same density profile after the perturbation allows us to follow the evolution of a large number of AMCs under many perturbations in a computationally feasible way. However, we note that this assumption is conservative. From Eq.~\eqref{eq:DeltaE_R}, energy is injected into and mass is lost predominantly from the outer parts of the AMC. The remnant should therefore be more dense after the perturbation, making it more resistant to further perturbations. Our assumption therefore leads to a smaller number of surviving AMCs than a more detailed (but computationally expensive) calculation.

In Fig.~\ref{fig:Perturbations}, we illustrate the evolution of the mass, radius, and binding energy of a minicluster under repeated perturbations. We fix the size of each perturbation to be 1000 times smaller than the initial binding energy, $\Delta E = 10^{-3} E_{\mathrm{bind}, i}$. In the PL case (left panel), we see that with each perturbation, mass is lost but energy is also injected and as a result the AMC becomes larger in size and is eventually disrupted entirely. This typically takes place with fewer encounters than expected without accounting for the evolution of the minicluster properties (which would be $\mathcal{O}(10^3)$ in this case, due to the fixed size of the perturbations). Miniclusters with NFW profiles (right panel) show a similar behaviour. Crucially, when $\Delta E$ becomes comparable to the remaining binding energy (after $\mathcal{O}(450)$ encounters in this example) the AMC radius begins to decrease. This is because for large $\Delta E/E_\mathrm{bind}$, almost all of the injected energy is carried away by the ejected mass ($f_\mathrm{ej}$ tends to one in  Fig.~\ref{fig:MassLoss}), leaving the remnant smaller and more dense than before the interaction. This behaviour --- seen also in studies of tidal shocks in globular clusters~\cite{2016MNRAS.463L.103G} --- emphasizes why we must carefully account for the redistribution of energy through Eq.~\eqref{eq:E_f}.

\section{Monte Carlo Simulations}
\label{sec:Simulations}

Having described how individual miniclusters are perturbed, we now want to understand the overall population of AMCs and their interactions over the history of the MW. We therefore run Monte Carlo simulations for this population of AMCs. For simplicity, we make a distinct split between structure formation and the stellar disruption of AMCs --- these two processes would typically happen concurrently (see \S~\ref{sec:caveats}).
Our simulations are therefore initialized with the expected properties of an AMC population today, at $z = 0$. We will discuss how removing this assumption could affect our results in \S~\ref{sec:caveats}. $N$-body simulations of a population of AMCs may be required to fully understand the details of the population today. Nevertheless, we attempt to capture the relevant physics in a simple and interpretable process.

We perform two different simulations using either the PL or NFW AMC density profiles. In reality, we expect the majority of AMCs to show an NFW profile for $M \gtrsim M_0$ and a PL profile for $M \lesssim M_0$, with the intermediate mass region being populated by both. 
Our simulations therefore attempt to bound the range of possible AMC populations today. We run the simulations for $t_\mathrm{MW} = 13.5\,$ billion years, therefore allowing for the maximum amount of disruption within the lifetime of the MW. The stellar distribution in the MW is assumed to be static in time and is modeled as a bulge plus a disk, as described in Appendix~\ref{sec:stellar_population}.
Again, we will discuss how changing these assumptions could change our results in \S~\ref{sec:caveats}.

\begin{figure}[tb!]
\includegraphics[width=0.49\textwidth]{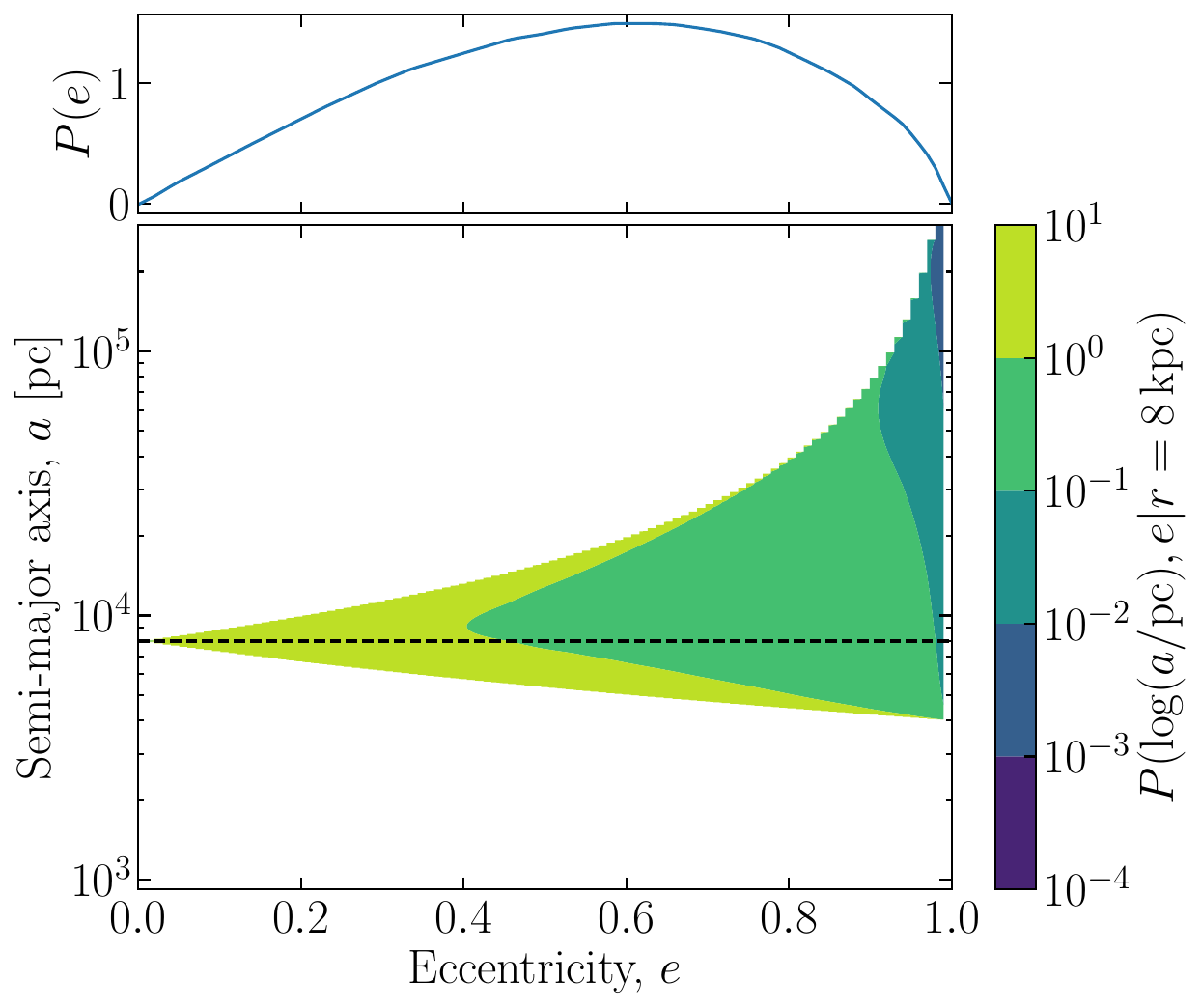} 
	\caption{\textit{Top} panel shows the eccentricity probability distribution for the orbits of AMCs taken from Ref.~\cite{vandenBosch:1998dc}. \textit{Bottom} panel shows the joint probability that an AMC will have a particular semi-major axis and eccentricity given a particular galactocentric radius $r=8\,\mathrm{kpc}$.} 
	\label{fig:P_a_e_given_r}
\end{figure}

\begin{figure}[t!]
\includegraphics[width=0.49\textwidth]{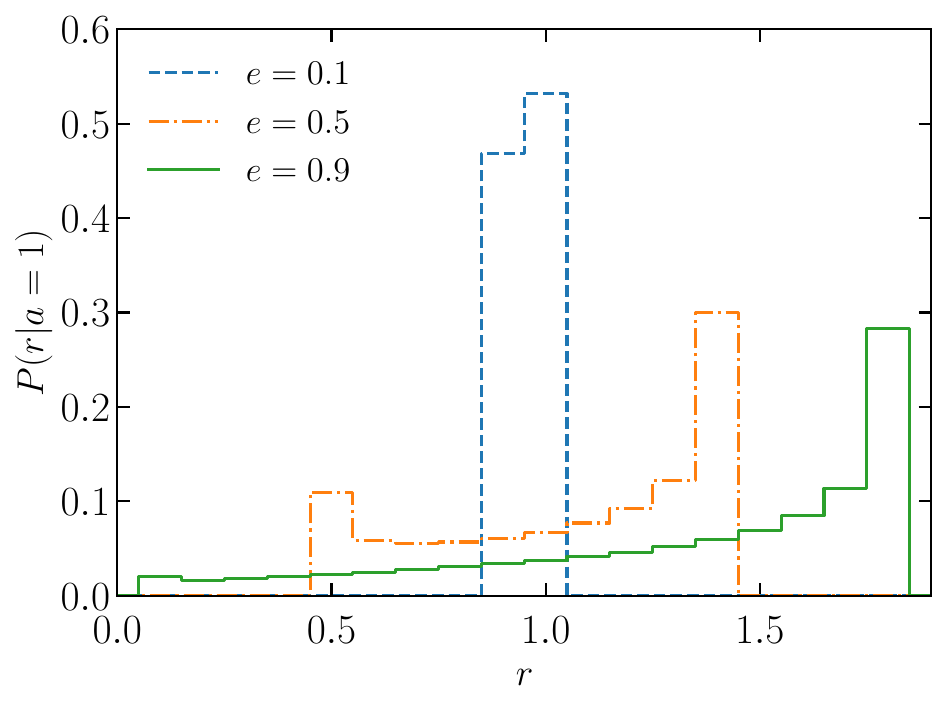}
	\caption{The binned probability of finding an AMC at a particular radius $r$ with a fixed semi-major axis $a=1$. We show distributions for eccentricity values: $e = 0.1$ (blue), $e = 0.5$ (orange), and $e = 0.9$ (green). }
	\label{fig:OrbitalRadius}
\end{figure}

Our simulated miniclusters are distributed in a spherically symmetric halo and follow elliptic orbits with a focus at the Galactic center, where eccentricities follow the distribution shown in the top panel of Fig.~\ref{fig:P_a_e_given_r}~\cite{vandenBosch:1998dc}. We show later in Fig.~\ref{fig:DensityReconstruction.pdf} that this reproduces the Galactic NFW density profile. For a given orbit, 
the time variation of the galactocentric radius $r$ of the orbit is described by the expression
\begin{equation}
    \label{eq:elliptic_drdt}
    \frac{\mathrm{d}r}{\mathrm{d}t} =  \sqrt{GM_{\mathrm{encl}}\left(\frac{2}{r} - \frac{1}{a} - \frac{a(1-e^2)}{r^2}\right)}\,,
\end{equation}
where $a$ and $e$ are the semi-major axis and the eccentricity of the elliptic orbit respectively. For simplicity, we take $M_{\mathrm{encl}}$ to be the mass of the MW enclosed within a sphere of radius $a$ given by the NFW profile~\cite{Navarro:1996gj}. The orbital radius is bounded by the values $a(1-e) \leq r \leq a(1+e)$ and the period is given by $T_\mathrm{orb} = 2\pi\sqrt{a^3/\(GM_{\mathrm{encl}}\)}$. The probability of finding an AMC at a specific radius at a particular instant in time is given by
\begin{equation}
    \label{eq:probability_elliptic}
    P(r|a,e) = \frac{2}{T_{\rm orb}}\,\(\frac{\mathrm{d}r}{\mathrm{d}t}\)^{-1}\,.
\end{equation}
In Fig.~\ref{fig:OrbitalRadius} we show the binned distribution of $P(r)$ for the values of the eccentricity $e = 0.1$ (blue), $e = 0.5$ (orange), and $e = 0.9$ (green). For small values of $e$, $P(r)$ approaches a delta function at $r = a$, i.e. a circular orbit. The values of $P(r)$ are larger at the boundaries $r = a(1\pm e)$ because the radial motion of the minicluster vanishes at either apsis and so $\mathrm{d}t/\mathrm{d}r$ diverges. Figure~\ref{fig:P_a_e_given_r} shows the joint probability distribution for $a$ and $e$ given a particular galactocentric radius (assuming the distribution for $e$ shown in the top panel of Fig.~\ref{fig:P_a_e_given_r}~\cite{vandenBosch:1998dc}). Importantly, it shows that the majority of AMCs found at a particular radius will have similar semi-major axes ($r \sim a$) and relatively low eccentricities $e\lesssim0.4$. At the same time, Fig.~\ref{fig:P_a_e_given_r} shows a non-negligible fraction of the population of AMCs found at a particular galactocentric radius will have highly eccentric orbits and a variety of semi-major axes.

Our Monte Carlo simulations range over a logarithmic grid of values for the semi-major axis $a \in [0.1, 50]\,\mathrm{kpc}$. For each value of $a$, the simulation proceeds as follows:
\begin{enumerate}
	\item We generate a set of $\NC = 10^5$ AMCs where the mass is sampled from a log-flat distribution between $M_{\rm min}$ and $M_{\rm max}$
	and in overdensity according to the distribution $\mathrm{d}f_{\rm AMC}/\mathrm{d}\delta$ in Eq.~\eqref{eq:dfddelta}. The log-flat sampling was used to ensure that we had a sufficiently large number of high mass AMCs in our simulations. We will discuss how to re-weight the results of the simulations (and how we apply the AS cut) to recover the true distribution in Sec.~\ref{sec:AMCtoday}. We draw the eccentricity of each AMC orbit from $P(e)$, which we treat as independent of the galactocentric radius. Finally, each AMC is also given a random inclination angle $\psi$ with respect to the Galactic plane, uniformly sampled within $[-\pi/2,\pi/2]$.
	\item For each AMC orbit, we compute the number density of the stellar field encountered by the AMC as a function of time $n_\star(t)$ over a full orbit (see Appendix~\ref{sec:stellar_population}). We then evaluate the total number of encounters with stellar objects over an orbit as
    \begin{equation}
        N = \int_0^{T_\mathrm{orb}} \, \mathrm{d}t\, n_\star(t) \, V_{\mathrm{AMC}}(t) \cdot \pi b_\mathrm{max}^2 \,,
        \label{eqn:Ninterac}
    \end{equation}
     where $b_\mathrm{max}$ is the maximum impact parameter that we consider (see below) and $V_{\mathrm{AMC}}(t)$ is the velocity of the AMC as a function of time. 
    The total number of interactions is then given as $N_{\mathrm{int}} = N \times(t_{\rm MW}/T_\mathrm{orb})$. We truncate the total number of interactions at $N_\mathrm{cut} = 10^6$ which we justify below. 
	\item For each AMC, we sample $N_{\mathrm{int}}$ relative velocities and impact parameters.
	From these we compute a list of energies $E_{\rm list}$ to be injected with each encounter, using Eq.~\eqref{eq:energy_perturbation}. The relative velocities are calculated by sampling from the integrand of Eq.~\eqref{eqn:Ninterac} to compute a distribution of the most likely interaction times. These interaction times can then be converted into a list of radii and AMC velocities using the Vis-Visa equation~\cite{logsdon1997orbital}. To obtain the encounter velocities, we then add a random 3-D velocity drawn from the local stellar velocity distribution, which we take as a Maxwell-Boltzmann distribution with dispersion $\sigma_v = \sqrt{G M_\mathrm{encl}(r)/r}$.
    The impact parameter is randomly drawn from the probability distribution defined as
    \begin{equation}
        \frac{\mathrm{d}P}{\mathrm{d}b} = \frac{2b}{b_\mathrm{max}^2}\,.
    \end{equation}
    We fix the maximum impact parameter $b_{\rm max}$ such that the energy injected is $1/N_\mathrm{cut}$ times smaller than the initial binding energy of the minicluster, $\Delta E (b_\mathrm{max}) = E_\mathrm{bind}/N_\mathrm{cut} = 10^{-6} E_\mathrm{bind}$.
    The truncation at $N_\mathrm{cut}$ is therefore not physically relevant for completely disrupted AMCs but is sufficiently large to capture the effects of partial disruption. We find typical values of $b_\mathrm{max} \sim 10^{-2} \,\mathrm{pc}$ and $b_\mathrm{max} \sim 10^{-1} \,\mathrm{pc}$ for PL and NFW profiles respectively.
	\item We then iteratively perturb each AMC, through $E_{\rm list}$, using the prescription described in \S~\ref{sec:interactions}. We recompute the new radius and mass after each iteration. As we note in \S~\ref{sec:Perturbation}, in some cases the density of an NFW minicluster can increase after an encounter, making it more resistant to further disruption. When this happens, we recompute $b_\mathrm{max}$ using the procedure described in the previous step. We then recompute the number of interactions in the \textit{remaining} simulation time and truncate this at $N_\mathrm{cut}$.
	If the total AMC energy, given in Eq.~\eqref{eq:AMCenergy1}, climbs above zero, we consider it to be completely disrupted and remove it from the simulation. Note that we do not keep track of which AMCs pass the AS cut \textit{during} the simulations, instead applying the AS cut to the distribution of perturbed AMCs, as described in \S~\ref{sec:PhysicalProperties}.
\end{enumerate}

\begin{figure}[tbh!]
\centering
\includegraphics[width=0.49\textwidth]{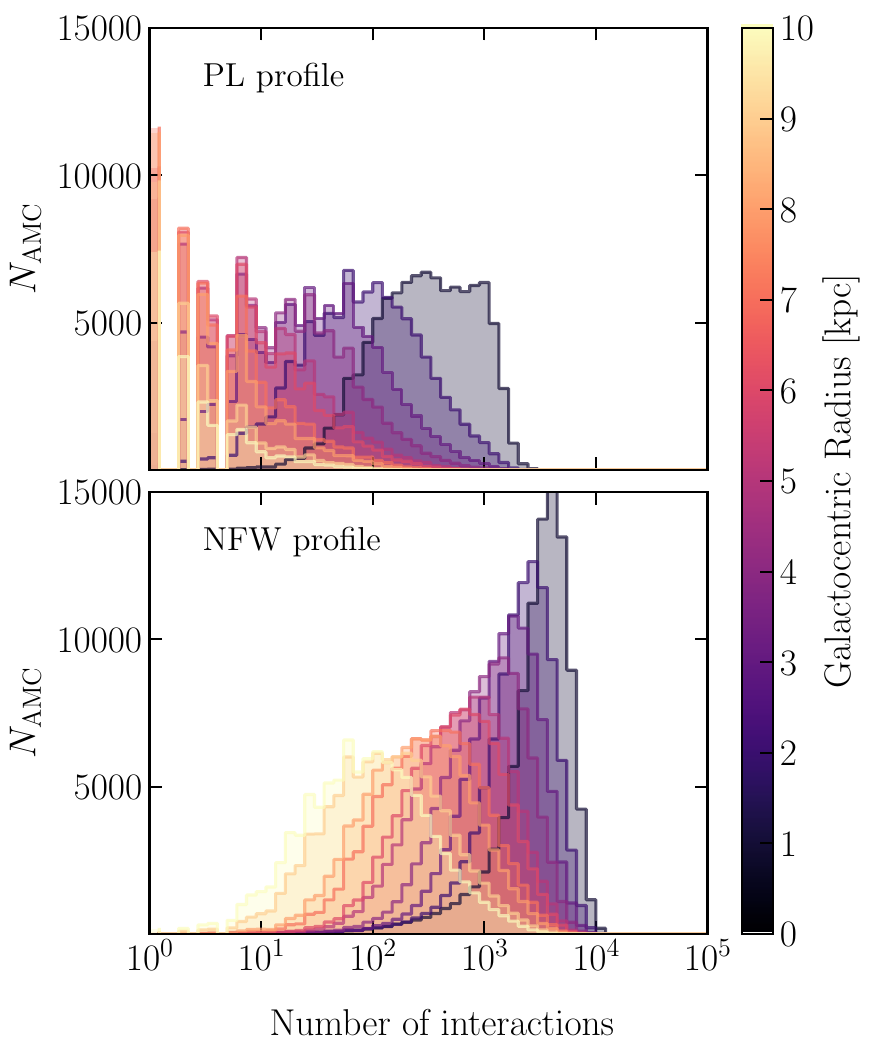}
\hspace*{-0.98cm}  
\includegraphics[width=0.39\textwidth]{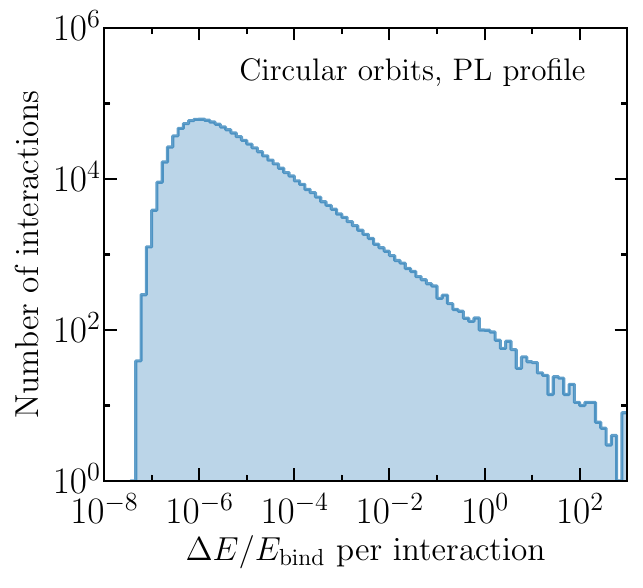}
	\caption{The \textit{top} and \textit{central} panels show the number of stellar interactions with AMCs on circular orbits at various galactocentric radii. Compared to AMCs with PL density profiles, miniclusters with NFW density profiles undergo significantly more interactions with $\Delta E/E_\mathrm{bind}$ at a fixed galactocentric radius. For illustration, in the \textit{bottom} panel we show the distribution of injected energies for PL profiles (which is approximately independent of galactocentric radii).}
	\label{fig:Ninteractions}
\end{figure}

Histograms of the number of interactions as a function of the galactocentric radius can be seen in the upper two panels of Fig.~\ref{fig:Ninteractions} --- to simplify the discussion, we show results for AMCs on circular orbits only. Note that here we count the number of simulated interactions for each AMC, stopping either at the end of the simulation time or when the AMC is disrupted.
There is a clear difference between the PL (top) and NFW (middle) simulations which can be seen as a distinct shift to larger numbers of interactions for NFW profiles for a fixed galactocentric radius. The shift originates from the smaller average density of the NFW miniclusters, which is reflected as a lower average binding energy. 
The reduced binding energy leads to more interactions above the threshold of $\Delta E/E_{\mathrm{bind}} > 10^{-6}$. In the NFW case, we also see that the number of interactions rarely extends beyond $10^4$. This is because the AMCs are either completely disrupted or are stripped to leave a high density remnant (for which further interactions above the threshold of $\Delta E/E_{\mathrm{bind}} > 10^{-6}$ are rare).

The lower panel of Fig.~\ref{fig:Ninteractions} shows the value of $\Delta E/E_\mathrm{bind}$ for a sample of $10^5$ interactions in the Monte Carlo simulation. The majority of interactions inject only a small amount of energy, while about 1 in 1000 interactions are strong enough to give rise to substantial mass loss $\Delta E \sim E_\mathrm{bind}$. This further justifies our cut on the maximum number of interactions $N_\mathrm{cut} = 10^6$, as this is much larger than the $\sim 1000$ interactions which would typically be required to unbind an AMC.

\subsection{Assumptions and Caveats}
\label{sec:caveats}

\textbf{Milky Way Properties ---} We have so far neglected the disruption of AMCs due to tidal stripping by the host halo of the Galaxy. The impact of tidal stripping can be quantified by considering the tidal radius of the AMCs, the distance from the center of the AMC at which tidal forces from the MW halo become comparable to the self-gravity of the bound AMC~\cite{vandenBosch:2017ynq}. We find that for PL profiles, the tidal radius is several orders of magnitude larger than the physical radius of the AMC, making them robust to tidal stripping. Instead, NFW profiles with $c = 100$ may be comparable in size to their tidal radius, especially at small Galactocentric radii. It is therefore likely that NFW AMCs have undergone some tidal stripping by this mechanism, from their initial concentration of $c \sim 10^4$, to reach our assumed concentration of $c = 100$. We therefore apply a correction of $5-40\%$ to the initial mass of NFW AMCs, to account for this mass-loss. However, given the small size of these corrections, we conclude that complete disruption by this mechanism is unlikely and subdominant to stellar disruption (see also Ref.~\cite{vandenBosch:2017ynq}). Full details concerning tidal stripping due to the MW halo are given in Appendix~\ref{app:host_halo}.



The orbital motion of an individual AMC is influenced by dynamical friction exerted by the MW~\cite[p. 644]{BinneyTremain:2008}. The orbit of an AMC with virial velocity $v(r)$ at the galactocentric radius $r$ decays with a timescale
%
\begin{equation}
    \label{eq:dynfrict}
    t_{\rm frc}^{-1} = \[4\pi G\rho(r)\]\,\frac{GM}{v^3(r)}\,\xi(r)\,,
\end{equation}
where $\xi(r) \sim \mathcal{O}\(10\)$ is a dimensionless function and $\rho(r)$ is the background density of the MW at the orbital radius $r$. Setting $\rho(r)$ to the Galactic NFW distribution, with the corresponding expression for the virial velocity, Eq.~\eqref{eq:dynfrict} gives $t_{\rm frc} \gtrsim t_{\rm MW}$ for $M \lesssim 10^{-5}\,M_\odot$ and $r \gtrsim 0.1\,$pc. Conversely, the orbits of the heaviest AMCs would be destabilized at very small Galactocentric radii. However, as we will see, tidal disruption by objects in the stellar bulge would disrupt these AMCs well before $t_{\rm MW}$. We can therefore ignore the effect of orbital decay.

Throughout this work, we only include tidal interactions with stars. In particular, we account for $\sim 10^{11}$ stars and fix their mass to be $1\,M_\odot$. Since the stellar mass function is relatively steep,\footnote{The Salpeter initial mass function, which is used almost universally, is given by $\mathrm{d}N/\mathrm{d}M \propto M^{-2.35}$~\cite{1955ApJ...121..161S}.} the vast majority of stars are around $1\,M_\odot$. We therefore expect that considering different stellar masses will produce only a small correction to our results. In addition, we have not considered tidal interactions with a separate population of NSs or white dwarfs which, despite having a mass of the same order as that of a typical star, are \textit{at least} an order of magnitude less numerous~\cite{2007coaw.book.....C}. Again we expect the corrections to our simulation results to be small when accounting for these additional astrophysical objects.

Finally, we neglect variations in the stellar density over the lifetime of the MW. Since the lifetime of a Solar mass star is $\mathcal{O}(10^{10})\,$years, the majority of stars born early in the MW's history will have finished their life cycle by today. It is therefore important to understand whether changes to the stellar abundance could affect our results. Luckily, the star formation rate is much larger than the death rate~\cite{Diehl:2006cf}, meaning that the stellar density has been increasing throughout the lifetime of the Galaxy. By using a stellar density as measured today we are therefore overestimating the amount of tidal stripping that could happen over the lifetime of the Galaxy. Nevertheless, the total stellar mass of galaxies like the MW is thought to be relatively constant (within a factor of two) since $z\sim1$~\cite{2007ApJ...665..265F}. Future work should adopt a time varying model of the MW which follows the cosmological star formation history~\cite{Madau:2014bja}.

\vskip 3pt

\textbf{Structure Formation ---} As mentioned above, for computational simplicity we made a distinct split between the hierarchical structure formation that these AMCs will undergo and their tidal stripping through interactions with stars. The interplay of these two physical effects requires a detailed study which we leave to future work. Nevertheless, our split represents a conservative approach since we allow for the \textit{maximum} amount of tidal stripping to occur for all AMC masses. Lighter AMC's are more abundant and are likely to have experienced fewer merger events than their heavier counterparts. These lighter AMC's have therefore been present in our Galaxy for the longest period of time --- our procedure should therefore be a good reflection of the tidal stripping for lighter AMCs. Heavier miniclusters, on the other hand, are less abundant and have been gradually merging throughout the history of our Galaxy. Mergers gradually increase the maximum mass that an AMC can achieve. For the heaviest AMCs, our simulations overestimate the amount of tidal stripping that may have occurred by today. On the other hand, Figs.~5 and 7 of Ref.~\cite{Fairbairn:2017sil} show that AMCs with $M \lesssim 10^5 \,M_0 \approx 10^{-6}\,M_\odot$ collapsed before $z \approx 10$ and therefore substantially before the formation of the MW. For these lower masses our simulations should capture the effects of stellar tidal interactions very well throughout the MW halo. 

For the most massive AMCs, from $10^{-6}\,M_\odot$ up to $M_\mathrm{max} \approx 5 \times 10^{-5}\,M_\odot$, there is still some uncertainty. At smaller galactocentric radii the survival probability is low ($r\lesssim 10\,\mathrm{kpc}$ for NFW profiles and $r\lesssim 3\,\mathrm{kpc}$ for PL profiles). For high mass AMCs (which formed through mergers) in these regions, we neglect the effects of concurrent structure formation and stellar interactions, leaving this to future work. In the outskirts of the MW halo $r\gtrsim 30\,\mathrm{kpc}$, stellar encounters are quite rare and will therefore not dramatically affect the growth of more massive AMCs. Fortunately, the high mass AMCs in the inner regions of the MW are a tiny proportion of the total number of miniclusters ($\mathcal{O}\left(10^{-9}\right)$) so we therefore conclude that our simulations are accurate for the majority of the AMC population.

\vskip 3pt

\textbf{Mutual AMC collisions ---} Throughout our simulations, we do not consider the mutual interactions between miniclusters. To see that this is a good approximation, we estimate the encounter rate of two AMCs at the Galactocentric radius $r$ as $\Gamma \sim n(r)v(r)\langle R^2\rangle$ where $n(r) \sim \rho_{\rm MW}(r)/\langle M_{\rm AMC}\rangle$ and $v(r)$ is the virial velocity associated with the NFW profile. These encounters occur rather frequently, for instance at $r \approx 4\,$kpc the encounter rate is $\Gamma\approx \(10^5{\rm\,years}\)^{-1}$. For comparison, the same computation for the encounter of a minicluster with a star yields $\Gamma \approx \(10^{19}{\rm\,years}\)^{-1}$.
However, mutual AMC encounters only deposit a small amount of energy during an interaction --- this can be seen from Eq.~\eqref{eq:energy_perturbation} which scales as the square of the mass of the perturbing object. The mean mass of an AMC is $\sim10^{-14} \,M_{\odot}$, meaning that a typical AMC-AMC interaction will deposit $10^{-28}$ times less energy than a typical AMC interaction with a star. Therefore, despite their large interaction rate, mutual AMC encounters will not significantly contribute to the tidal disruption of miniclusters.

In addition to AMC collisions that lead to tidal disruption, mutual miniclusters interactions can also lead to mergers. Importantly, the increased background density within the MW with respect to the critical density will cause these merger interactions to happen more regularly than in typical simulations. Fortunately, this effect will be most prominent in the Galactic center where the density is largest, but also where stellar disruption will be dominant. We leave a complete study of this effect to future work.

\vskip 3pt

\textbf{Minicluster Substructure ---} As discussed in Sec.~\ref{sec:PhaseSpace}, we expect the most massive AMCs to form from hierarchical mergers of lighter ones. Numerical simulations of AMC clustering show that an internal granular structure is expected on top of an overall NFW distribution~\cite{Eggemeier:2019khm}. We have not considered such a granular substructure, since lighter AMCs are expected to be dissolved within the larger minicluster over the lifetime of the MW due to the effects of dynamical friction~\cite{Dai:2019lud}.

\section{Axion Miniclusters Today}
\label{sec:AMCtoday}


\begin{figure*}[tbh!]
\includegraphics[width=1.\textwidth]{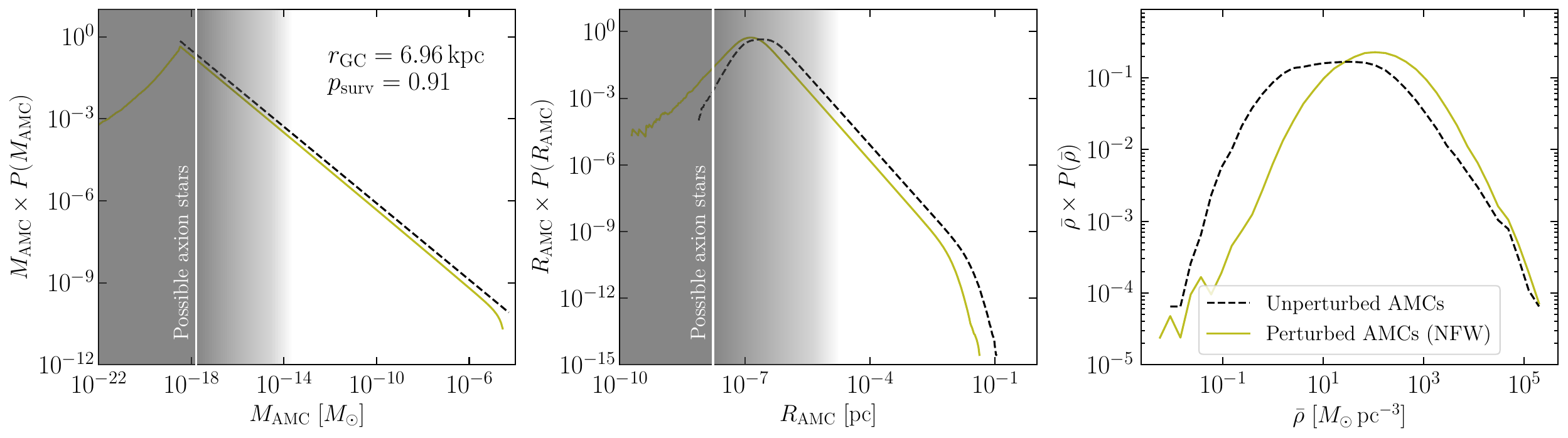}
	\caption{Example of AMC properties from our Monte Carlo simulations before (black dashed) and after (solid olive) disruption. We show the probability distributions of the mass $M_\mathrm{AMC}$, radius $R_\mathrm{AMC}$, and  mean density $\bar{\rho}$. We assume these to have NFW internal density profiles and to be on circular orbits with a galactocentric radius of $r = 6.96 \,\mathrm{kpc}$, leading to a survival probability $p_\mathrm{surv} = 0.91$. These probability distributions are obtained through the reconstruction procedure described in \S~\ref{sec:PhysicalProperties}. The grey shading indicates the regions of the parameter space that are removed by the AS cut. The white lines shows the smallest value of the corresponding parameter that passes the AS cut.
	}
	\label{fig:histdensity}
\end{figure*}

In this section we first discuss how to construct the \textit{true} AMC population distributions today from our Monte Carlo simulations followed by a discussion of the results. Note that although we present results for a specific choice of HMF, the reconstruction procedure allows us to use the same Monte Carlo results for arbitrary HMFs. In particular, this could include changes in the mass cut-offs $M_\mathrm{min}$ and $M_\mathrm{max}$, changes to the AS cut, or the introduction of correlations between the minicluster mass  $M_\mathrm{AMC}$ and overdensities $\delta$. The full suite of results and the corresponding code to reinterpret them can be found at \href{https://github.com/bradkav/axion-miniclusters/}{github.com/bradkav/axion-miniclusters} \cite{AMC_code}.

\subsection{Reconstructing Physical Properties}
\label{sec:PhysicalProperties}

Each of our Monte Carlo samples corresponds to $N = 10^5$ AMCs with orbits of a given semi-major axis $a$. In order to calculate the survival probability and AMC properties as a function of galactocentric radius $r$, we must assign each sample a weight $w$, proportional to the time that AMC spends at a given value of $r$, from Eq.~\eqref{eq:probability_elliptic}. With this, we essentially `smear' each AMC sample over a range of radii $r$, allowing us to estimate the properties as a function of $r$ (instead of $a$). We assume that the initial number density of miniclusters \textit{as a function of semi-major axis} $n_\mathrm{AMC}(a)$ follows an NFW profile. For a single AMC with semi-major axis $a_i$ and eccentricity $e_i$, the weight assigned in some radial bin $\Delta r$ is then:
\begin{align}
    \begin{split}
        w = \left[\frac{\Delta a_i}{N}\right] \times \left[4 \pi a_i^2 n_\mathrm{AMC}(a_i)\right] \times \left\langle P(r|a_i, e_i) \right\rangle_{\Delta r}\,.
    \end{split}
\end{align}
The first term accounts for the fact that the Monte Carlo samples are not uniformly distributed in $a$, but concentrated on logarithmically-spaced grid points, with spacing $\Delta a_{i}$. The second term is the assumed initial probability distribution for the semi-major axis $P(a)$, with $n_\mathrm{AMC}$ defined in Eq.~\eqref{eq:MCdistribution}. The final term is the fraction of time spent at a given radius, defined in Eq.~\eqref{eq:probability_elliptic}, averaged over the radial bin of interest. The AMC number density at a given Galactocentric radius can be obtained by summing over the weights of all AMCs (with potential contributions from all values of $a$). This smearing procedure gives rise to an approximately NFW profile \textit{as a function of galactocentric radius} $n_\mathrm{AMC}(r)$, as shown by the black dashed line in Fig.~\ref{fig:DensityReconstruction.pdf}.

The number density of AMCs at a galactocentric radius $r$ can be written as:
\begin{equation}
    \frac{\mathrm{d}n}{\mathrm{d}M\mathrm{d}R}(r) = p_{\rm surv}(r)\,\nC(r)\,P(M, R|r)\,,
    \label{eq:define_psurv}
\end{equation}
where the survival probability is defined as the ratio of the number of surviving AMCs $N_{\rm surv}(r)$ to the number $N_{\rm initial}(r)$ at a given radius: $p_{\rm surv}(r) = N_{\rm surv}(r)/N_{\rm initial}(r)$.
The probability distribution for the minicluster mass and radius at a given galactocentric radius $P(M, R|r)$ can be extracted from the Monte Carlo simulations.\footnote{For clarity, we now drop the subscript $\mathrm{AMC}$ from $M_\mathrm{AMC}$, $R_\mathrm{AMC}$, etc.}

Our Monte Carlo simulations were performed with a log-flat distribution of AMC masses. The results must then be adapted to reflect the true mass function of the AMCs. From our simulations, we obtain the final density $\rho$ and the mass-loss fraction $\nu = M_f/M_i$ of each AMC in a sample.\footnote{For AMCs with PL density profiles, there is no mass loss, so $\nu_k = 1$ for all AMCs.} The simulations confirm that the distribution of $\rho$ and $\nu$ do not depend on the initial AMC mass but only on the initial density, as discussed in \S~\ref{sec:interactions}. 

For a given initial mass function $P_i(M_i)$, the final distribution of masses can be obtained by integrating over all possible initial masses $M_i$, selecting only those which produce the correct final mass $M_f$.
The joint distribution of AMC mass and density  after disruption can then be written as:
\begin{align}
\begin{split}
    P(M_f, \rho|r) &= \int \delta(M_f - \nu M_i) P_i(M_i) P(\rho, \nu|r)\,\mathrm{d}\nu\,\mathrm{d}M_i \,\\
    &=
    \int \frac{1}{\nu}P_i(M_f/\nu) P(\rho, \nu|r)\,\mathrm{d}\nu\,.
\end{split}
\end{align}
Here, $P(\rho, \nu|r)$ is the final probability distribution for the density and mass-loss fraction at a given galactocentric radius. We can now write the final mass function as a Monte Carlo integral:
\begin{align}
\begin{split}
\label{eq:P_M_MonteCarlo}
    P(M_f|r)  &= \int\int \frac{1}{\nu}P_i(M_f/\nu) P(\rho, \nu|r)\,\mathrm{d}\nu\,\mathrm{d}\rho\\
    &\approx \sum_k \frac{w_k}{\nu_k} P_i(M_f/\nu_k)\,,
\end{split}
\end{align}
where we have replaced the integral by a sum over the $N$ surviving AMCs, with properties  $(\rho_k, \nu_k)$, distributed according to $P(\rho, \nu)$. 
The index $k$ runs over the AMCs in the sample, with the weights $w_k$ calculated at a given galactocentric radius, as described above. Here, AMCs which have been completely disrupted are excluded from the integral (or the corresponding sum). The distribution of final radii $R_f$ can also be written as:
\begin{align}
\begin{split}
\label{eq:P_R_MonteCarlo}
    P(R_f|r) \approx \left. \sum_k \frac{w_k}{\nu_k} \left(\frac{3 M_f}{R_f}\right) \,P_i(M_f/\nu_k)\right|_{M_f = \frac{4\pi}{3} \rho_k R_f{}^3}\,,
\end{split}
\end{align}
and similarly for any other distribution of interest. We fix the initial mass function according to Eq.~\eqref{eq:mcdistribution}, with slope $\gamma = -0.7$. For NFW AMCs, we also apply a correction of $5-40\%$ to the initial AMC mass to account for tidal stripping due to the MW host halo. This is implemented directly in the definition of $P_i(M)$ for NFW AMCs. Full details are given in Appendix~\ref{app:host_halo}.

Since the disruption process is mainly sensitive to the densities of the AMCs (and not the AMC masses), we do not expect a significant difference to our results when changing the HMF. For consistency, we therefore re-run the entire pipeline using a different slope $\gamma = -0.5$ and find only a $\sim 10\%$ increase in the survival probability (compared to $\gamma = -0.7$) for NFW AMCs at the Solar position. This is not due to a change in the disruption properties of the AMCs but rather a change in the fraction of AMCs passing the final AS cut. There is no appreciable change for PL miniclusters. We therefore do not consider $\gamma = -0.5$ further.

In our numerical results, we assume that all axions are bound in AMCs ($f_\mathrm{AMC} = 1$) with masses between $M_\mathrm{min}$ and $M_\mathrm{max}$, given in Eq.~\eqref{eq:Mminmax}. For our `AS cut' results we use the \textit{total} perturbed sample of AMCs and require that the AS radius be smaller than the AMC radius. This reduced sample is then compared to the unperturbed sample with the same cut applied (as described in \S~\ref{sec:ASs}). As mentioned in \S~\ref{sec:ASs}, this cut effectively reduces the fraction of axions bound in AMCs.

The AS radius in Eq.~\eqref{eq:axionstarradius} can be re-written as
\begin{equation}
    R_\mathrm{AS} =  R_\star \left( \frac{M_\mathrm{AMC}}{M_\star}\right)^{-1/3}\,,
\end{equation}
where we define the constants $R_\star = 1.7 \times 10^{-6} \,\mathrm{pc}$ and $M_\star = 10^{-16}\,M_\odot$.
The AS cut therefore requires that
\begin{equation}
    R_f > R_\mathrm{AS}(M_i) = R_\star \left( \frac{M_i}{M_\star}\right)^{-1/3}\,,
\end{equation}
where we calculate the AS radius using the initial AMC mass, assuming that the properties of the central AS are unaffected by perturbations.
In Eq.~\eqref{eq:P_M_MonteCarlo} and Eq.~\eqref{eq:P_R_MonteCarlo}, this cut is equivalent to summing only over those samples which satisfy
\begin{equation}
    \rho \leq \frac{1}{\nu} \frac{3 M_\star}{4 \pi R_\star^3} \left(  \frac{M_f}{M_\star}\right)^2 \approx 4.66 \,M_\odot \,\mathrm{pc}^{-3}\, \frac{1}{\nu} \left(  \frac{M_f}{M_\star}\right)^2\,.
\end{equation}

\subsection{Results}

\begin{figure*}[t!]
\centering
\includegraphics[width=0.49\textwidth]{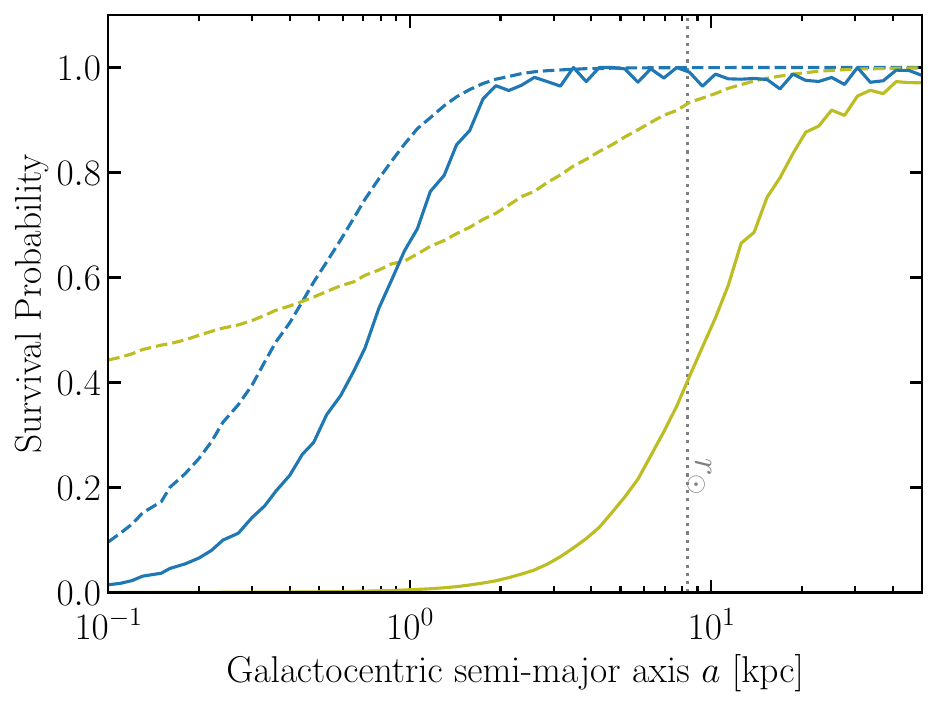} \includegraphics[width=0.49\textwidth]{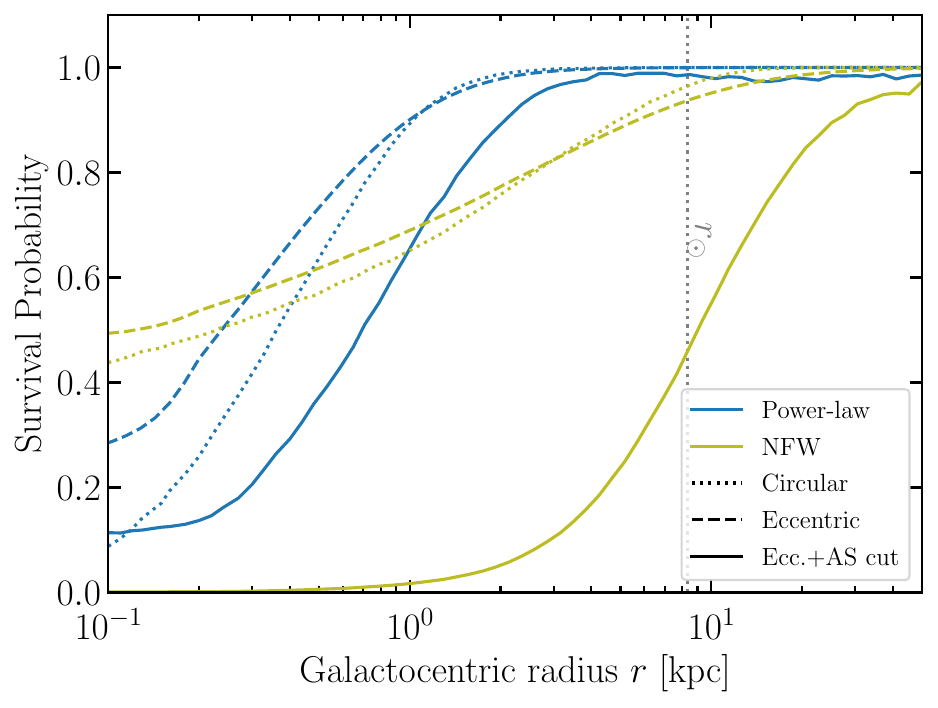}
	\caption{Survival probability of AMCs as a function of their galactocentric semi-major axes (left) and galactocentric radii (right). The vertical dashed line marks the position of the Solar System. In the \textit{right} panel we show the survival probability for both circular (dotted) and eccentric (dashed) orbits. The eccentric orbits are \textit{smeared out} according to the proportion of time they spend at a given galactocentric radius (as described in \S~\ref{sec:PhysicalProperties}). Finally, we show the survival probability for miniclusters with eccentric orbits that pass our AS cut (solid line) where we have normalized this to be one at large galactocentric radii.
}
	\label{fig:survival}
\end{figure*}

In Fig.~\ref{fig:histdensity}, we show an example of the reconstructed probability distributions of AMC properties at the end of our Monte Carlo simulations. Specifically, we show results for AMCs with NFW density profiles and, for simplicity, on circular orbits with Galactocentric radius of $r = 6.96\,\mathrm{kpc}$. In this case, only around 9\% of miniclusters are destroyed --- this can be seen in Fig.~\ref{fig:survival} from the olive dashed line --- but the properties of those which survive are drastically altered. The results in Fig.~\ref{fig:histdensity} do not include the AS cut. However, the shaded areas indicate regions of the parameter space which are progressively removed by this cut. In particular, no AMCs pass the AS cut to the left of the vertical white lines.

Focusing on the left panel of Fig.~\ref{fig:histdensity}, we see that the low-mass tail of the distribution of AMC masses extends to lower values after disruption is taken into account. This is due to mass loss from miniclusters which are perturbed but not completely disrupted (see \S~\ref{sec:Perturbation}). In some cases, the final mass of an AMC is reduced by several orders of magnitude compared to its initial mass. We have verified also that the fractional mass loss \textit{does not} depend on the initial mass of the minicluster but only on its density. 

The right panel in Fig.~\ref{fig:histdensity} demonstrates the strong dependence of the disruption on the initial AMC density. From our discussion in Sec.~\ref{sec:stripping}, we expect the amount of energy injected per encounter to scale as $\Delta E/E_\mathrm{bind} \sim R_\mathrm{AMC}^3/M_\mathrm{AMC} \sim 1/\rho_\mathrm{AMC}$. Indeed, we see that very dense miniclusters undergo little disruption. Instead, less dense miniclusters (e.g.\ around $1 \,M_\odot \,\mathrm{pc}^{-3}$) have a low survival probability and those which survive lose mass. If the AMCs undergo only small perturbations, they may increase in radius through energy injection, leading to a tail of diffuse miniclusters (down to $\sim 10^{-3}\,M_\odot \,\mathrm{pc}^{-3}$). Instead, large perturbations can cause substantial mass loss from the AMCs, but very little energy is injected into the remnant. This leads to an overall increase in the typical AMC density. The distribution of AMC radii (central panel in Fig.~\ref{fig:histdensity}) follows from this same argument. The typical AMC is more dense after accounting for stellar interactions, leading to a reduction in the AMC radius.

\vskip 3pt

In Fig.~\ref{fig:survival}, we show the survival probability of our simulated miniclusters as a function of both their semi-major axes (left) and galactocentric radii (right) for both PL and NFW internal density profiles. The right panel shows results for both eccentric and circular orbits where the former is constructed using the prescription described in \S~\ref{sec:PhysicalProperties}. In both panels we see that there is a characteristic \textit{transition} from the high stellar density inner region $r \lesssim \mathcal{O}(1) \,\mathrm{kpc}$ (where the number of interactions between stars and AMCs is so high that almost all miniclusters are completely disrupted) to a low stellar density outer region $r \gtrsim \mathcal{O}(10) \,\mathrm{kpc}$ (where interactions are rare). This change in the number of interactions can be seen clearly in Fig.~\ref{fig:Ninteractions}. Focusing on the left panel of Fig.~\ref{fig:survival} we see a distinct shift of the transition region to larger radii from the PL to NFW density profiles. This shift comes from the enhanced number of interactions at a fixed galactocentric radius for NFW miniclusters --- this can also be seen in Fig.~\ref{fig:Ninteractions}.

In the right panel of Fig.~\ref{fig:survival}, we see this same distinct shift from PL to NFW profiles for both eccentric and circular orbits. Moving from circular to eccentric orbits produces a \textit{smearing} of the transition region caused by the distributions of semi-major axes and eccentricities that contribute to the AMCs at a particular galactocentric radius, as shown in Fig.~\ref{fig:P_a_e_given_r}. For example, at low galactocentric radii, the AMC density will have contributions from both quasi-circular orbits which spend a large amount of time in dense stellar regions and highly eccentric orbits that spend the majority of their time in low density stellar environments at larger radii.

We also show results for the survival probability of AMCs which pass the AS cut (solid lines). In this case, we normalize the survival probability to one at large galactocentric radii, which is equivalent to factoring out the initial fraction of AMCs which pass the cut,
$f_\mathrm{cut}^\mathrm{PL} = 2.7 \times 10^{-4}$ for PL density profiles and $f_\mathrm{cut}^\mathrm{NFW} = 1.5 \times 10^{-2}$ for NFW profiles. The survival probability with the AS cut is always smaller than without the cut. This is because stellar perturbations can strip the AMCs until their radius drops below the corresponding AS radius. This demonstrates that the AMC properties can be substantially altered by stellar interactions. At the Solar radius, we find a survival probability (including the AS cut) of 99\% for AMCs with PL profiles and 46\% for AMCs with NFW profiles.

\begin{figure}[tbh!]
\includegraphics[width=0.49\textwidth]{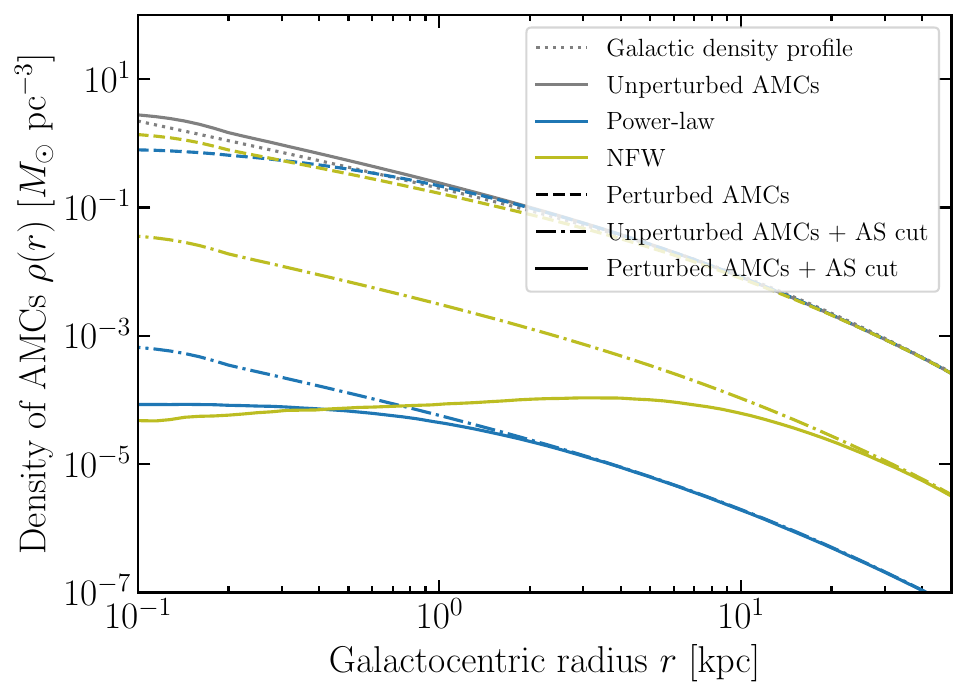}
	\caption{Density of simulated AMCs before taking into account stellar interactions (solid gray), compared with the expected NFW profile for the DM halo (dotted gray). The spatial distribution of AMCs with PL and NFW internal density profiles are shown as blue and olive lines respectively. The density profiles are extracted from the Monte Carlo simulations according to the weighting procedure in Sec.~\ref{sec:AMCtoday}. We also show the distribution of unperturbed and perturbed AMCs passing the AS cut (dot-dashed and solid lines respectively).}
	\label{fig:DensityReconstruction.pdf}
\end{figure}

Finally, Fig.~\ref{fig:DensityReconstruction.pdf} shows the density of AMCs as a function of the galactocentric radius. Firstly, we show that we are able to correctly reconstruct the Galactic NFW profile (gray dotted line) using the initial sample of unperturbed miniclusters with NFW-distributed semi-major axes (gray solid line). The fraction of AMCs that are removed through the AS cut is indicated by the reduced normalization of the dot-dashed lines. The density of perturbed miniclusters passing the AS cut is shown by the solid lines. The PL profile AMCs (blue solid line) maintain a population down to small radii --- these AMCs are still likely to have undergone many interactions and therefore may have significantly different properties to those at the start of the simulation. In contrast, the NFW miniclusters (olive solid line) show a sharp reduction in density at around $r \sim 10\,\mathrm{kpc}$.

\section{Applications}
\label{sec:apps}
Many experimental probes of axion DM require one to make assumptions about both the large scale halo distribution of DM and its structure on small scales. 

Here we briefly investigate the phenomenological consequences of our results for a few primary observational channels.

\subsection{Lensing}

Gravitational lensing has been used to constrain the fraction of DM in the form of faint compact objects~\cite{Paczynski:1985jf}.
Surveys actively search for microlensing events caused by objects passing through the line of sight between the Earth and stars in target structures like the Magellanic System~\cite{Tisserand:2006zx, Alcock:2000ph, 2015AcA....65....1U, Niikura:2019kqi,Croon:2020wpr}, the inner Galactic bulge~\cite{2015AcA....65....1U, Niikura:2019kqi,Croon:2020wpr}, and M31~\cite{Niikura:2017zjd,Croon:2020ouk}.

For simplicity, we limit our discussion to microlensing from a point-like lens of mass $M$ with a point-like background source --- in reality both the lensing AMC and the background galaxy are extended objects. To assess the impact of our Monte Carlo results we compute the difference between the expected number of lensed events, perturbed and unperturbed, for a variety of observational targets. We leave a more detailed analysis for future work.

We denote the observer-source, observer-lens, and lens-source distances as $D_{\rm S}$, $D_{\rm L} \equiv x\,D_{\rm S}$, and $D_{\rm LS} = \(1-x\)\,D_{\rm S}$ respectively with $0 \leq x \leq 1$. We expect the set of microlensing events over some observation time to be Poisson distributed with the expected number of events given by~\cite{Alcock:2000ph}
\begin{equation}
    \label{eq:expected_number}
    \bar N_{\rm ex} \propto \int \mathrm{d}t\,\int \mathrm{d}x\,\frac{\mathrm{d}^2\Gamma}{\mathrm{d}x\,\mathrm{d}t}\,.
\end{equation}
The differential event rate for a single source star with respect to distance and event time, $\mathrm{d}^2\Gamma/(\mathrm{d}x\,\mathrm{d}t)$, depends on the mass distribution of AMCs, their velocity distribution, and is proportional to the number density
of lenses along the line of sight $n_{\rm AMC}(x)$. Here we treat $n_{\rm AMC}(x)$ as a proxy for the expected number of lensing events.

We consider microlensing events from sources residing in the following targets: the MW Galactic bulge, M31, and the Large Magellanic Cloud (LMC). Miniclusters in either the halo of the MW or in the target's halo could lead to lensing events. We model the DM halo distribution in each target galaxy and in the MW according to the NFW density profile in Eq.~\eqref{eq:density_NFW}. In Table~\ref{tab1} we report the distances between the Solar System and the source $D_{\rm S}$, together with the Galactic longitude and the latitude $\(\ell_g, b_g\)$ and the parameters used to model the NFW profile. For the LMC and M31 we use the AMC survival probability obtained in the MW but rescaled by the scale factor $r_s$ of the target galaxy. 
The distribution of AMCs between the Solar System and the source is given by the sum of the profiles along the line of sight.

\begin{table}[tb!]
\def\arraystretch{1.5}
	\begin{tabular}{ccccc}
    \cline{2-5}
    \hline\hline
		& $D_{\rm S}\[{\rm kpc}\]$ & $(\ell_g, b_g)$ & $r_s\[{\rm kpc}\]$ & $\rho_s\[M_\odot/{\rm kpc}^3\]$ \\ 
		\hline
		MW bulge & 8.3 & $\(1.09^\circ, -2.39^\circ\)$ & $16.1$ & $11.8\times 10^6$ \\ 
		LMC & 48 & $\(280.5^\circ, -32.9^\circ\)$ & $12.6$ & $2.6\times 10^6$ \\
	    M31 & 780 & $\(121.2^\circ, -21.6^\circ\)$ & 25.0 & $5.0\times 10^6$ \\ \hline\hline
	\end{tabular}
	\caption{Parameters used for the source location and halo modelling. We report the distances of the LMC~\cite{Pietrzynski:2013gia} and M31. We also specify the NFW parameters used for the MW~\cite{Dehnen:2006cm}, for the LMC~\cite{Buckley:2015doa}, and for M31~\cite{Klypin:2001xu}.}
	\label{tab1}
\end{table}

\begin{figure}[tbh!]
 	\includegraphics[width=1\linewidth]{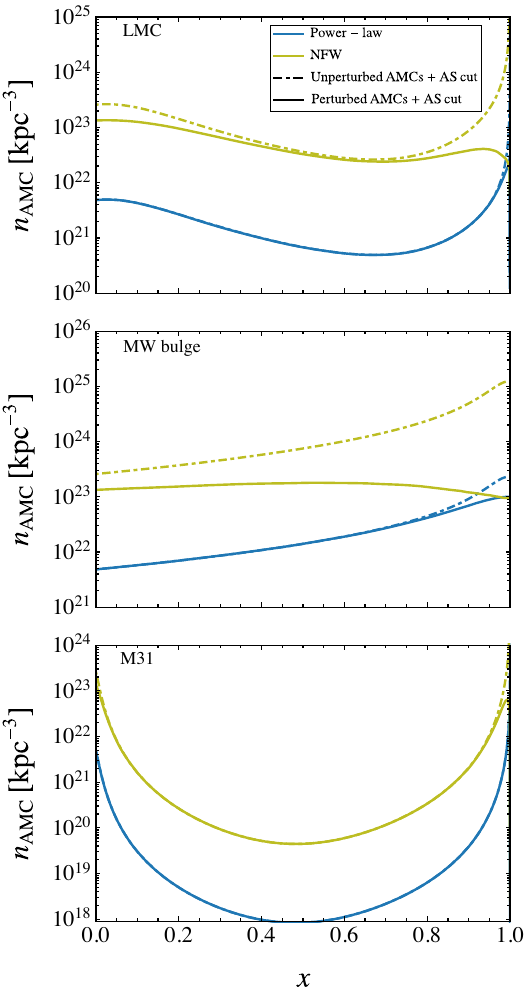}
 	\caption{The number density of AMCs (in kpc$^{-3}$) as a function of the distance of the lens $x = D_{\rm L}/D_{\rm S}$ for the unperturbed population (dot-dashed line) and for the total population of perturbed AMCs (solid lines).
 	We show the results for target sources placed in the LMC (top), in the MW bulge (middle), and in M31 (bottom).  For the both perturbed and unperturbed populations, we have applied to AS cut.}
 	\label{fig:lensing} 
\end{figure}

Results are shown in Fig.~\ref{fig:lensing} for the LMC (top), the MW bulge (middle), and M31 (bottom). We show the number density of AMCs as a function of the ratio of the lens distance to the source distance $x = D_{\rm L}/D_{\rm S}$ for the perturbed population derived from our simulations (solid line), assuming NFW (olive) and PL (blue) AMC density profiles, including the AS cut, and using the same color code as in Fig.~\ref{fig:survival}. For comparison, we also show the results when perturbations are neglected (dot-dashed line).

In Fig.~\ref{fig:lensing}, considering observations of the LMC (top panel), the number density of the NFW AMC population is strongly suppressed both towards the inner region of the MW ($x \ll 1$) and the LMC ($x \approx 1$), with respect to the unperturbed one. This behavior is evident also for AMCs along the line of sight towards the Galactic center (middle panel), for both PL and NFW profiles. For M31 (bottom panel) the difference between the populations is not clearly visible, except near the center of M31 where the number density is significantly affected. This is because we fix the line of sight to end at the center of the target galaxy, where the distribution of perturbed AMCs drops significantly. Fixing the line of sight to point at other regions of the target galaxies, where the survival probability of AMCs is higher, would lead to smaller differences. Note that for PL profile AMCs, the number density closely follows the unperturbed distribution for $x \ll 1$, i.e. within the MW. Compared to those with PL profiles, NFW miniclusters show a more dramatic reduction in the number density along the line of sight for all three sources. This is due to the difference in survival probabilities shown in Fig.~\ref{fig:survival}.

Using Eq.~\eqref{eq:expected_number}, we denote the fractional decrease in the expected number of lensing events before and after tidal interactions as $\delta_N \equiv \Delta \bar N_{\rm ex} / \bar N_{\rm ex}$. For sources in M31 we expect $\delta_N \approx 0.5\%$ (PL) or $\delta_N \approx 18\%$ (NFW), where the discrepancy is almost entirely due to the AMC disruption in the target galaxy. When considering the LMC as a source, we find $\delta_N \approx 1\%$ (PL) or $\delta_N \approx 32\%$ (NFW). The largest difference is clearly seen in searches towards the MW bulge where $\delta_N \approx 12\%$ (PL) or $\delta_N \approx 92\%$ (NFW). Even when the fractional decrease in the number of events is small, the properties of these events could change dramatically. For instance, the duration of the microlensing events can be significantly shortened since the surviving AMCs are each less massive than in the unperturbed case. A more careful analysis would be required to determine the predicted properties of AMC microlensing events.

\subsection{Direct Detection}

The main direct detection experiments that would be affected by the presence of AMCs  are `haloscopes'. Haloscopes convert cosmic axions into a detectable signal inside a resonant cavity immersed in a strong magnetic field~\cite{Sikivie:1983ip, Sikivie:1985yu}. Once the settings for the cavity have been fixed, the power output $P_a$ from the conversion of the cosmic axions in the magnetic field of the cavity is proportional to the local energy density of axions $\rho_a$ times $g_{a\gamma\gamma}^2$~\cite{Sikivie:1983ip, Sikivie:1985yu}. The value of the local energy density has contributions from a smooth component in the Solar neighborhood and from substructures like streams and AMCs. This means that for a fixed value of $g_{a\gamma\gamma}$, $P_a$ can still vary significantly. In particular, an AMC passing near the Earth would enhance the power output of the cavity for a short amount of time. This is expected to be a rather rare event, as estimates show that an encounter would occur only every $10^4-10^6$ years~\cite{Sikivie:2006ni, Visinelli:2018wza}. 

Here, we compute the encounter rate for AMCs of radius $\RC$ with the geometrical cross section of the Earth $\sigma = \pi \RC^2$. In principle, a correction due to gravitational focusing becomes important for miniclusters with radii smaller than $\bar R = 2G_N\,M_{\oplus}/\sigma_u^2 \approx 10\,$km, where $M_{\oplus}$ is the mass of the Earth and $\sigma_u \approx \mathcal{O}\left(200{\rm \,km/s}\right)$ is the velocity dispersion at the Earth's location.
Since miniclusters are much larger than $\bar R$, we neglect this focusing contribution. The encounter rate between the Earth and the population of miniclusters is
\begin{equation}
	\label{eq:Earthrate}
	\Gamma_{\odot} = \nC(r_\odot)\,\overline{\langle\sigma u \rangle}(r_\odot)\,,
\end{equation}
where $\nC(r_\odot)$ is the local number density of AMCs, $r_\odot = 8.33\,$kpc is the distance of the Solar System from the MW Galactic center~\cite{Gillessen:2008qv} and $\overline{\langle\sigma u \rangle}(r_\odot)$ is the velocity-averaged cross section weighted with the probability distribution of $\RC$ at $r_\odot$.

The local number density is given by Eq.~\eqref{eq:MCdistribution} times the fraction $f_{\rm cut}$ that accounts for the AS cut in the HMF given in \S~\ref{sec:ASs}. We compute $\overline{\langle\sigma u \rangle}(r_\odot)$ from the distribution of AMC radii at $r = r_\odot$ obtained from the Monte Carlo simulations for both the {\it unperturbed} and the {\it perturbed} populations, taking into account the full distribution of masses and densities. The number densities of perturbed AMCs are also weighted by the survival probability in Fig.~\ref{fig:survival}.
Two competing effects combine in the computation of the encounter rate for the perturbed population. In general, successive perturbations tend to \textit{puff up} AMCs, making their mean radius larger and encounters with Earth more probable. On the other hand, stellar encounters destroy some of the AMCs (or strip them to below the AS cut), thereby lowering the chance of encounters. For the PL profile the two effects compensate, leading to the same encounter rate $\Gamma_{\odot} \approx \(4\times 10^6 {\rm \,years}\)^{-1}$ for both the {\it unperturbed} and the {\it perturbed} AMC distributions. For the NFW profile, the first effect dominates, leading to $\Gamma_{\odot} \approx \(10^3 {\rm \,years}\)^{-1}$ for {\it unperturbed} AMCs and $\Gamma_{\odot} \approx \(4\times 10^3 {\rm \,years}\)^{-1}$ for {\it perturbed} AMCs. The difference in the magnitude of the results between the NFW and the PL populations is due to the larger fraction of NFW AMCs which pass the AS cut as well as the larger typical radius for NFW profiles. Recall that the results for $\Gamma_\odot$ are inversely proportional to the AMC fraction, which we set to $f_{\rm AMC} = 1$.

\begin{figure}[th!]
 	\includegraphics[width=1\linewidth]{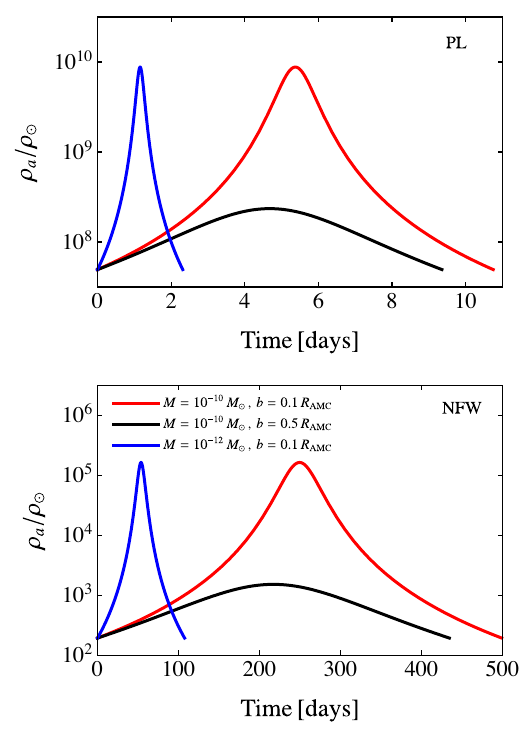}
 	\caption{The enhancement in the local axion density due to the encounter of Earth with an AMC, in units of $\rho_\odot = 0.45\,$GeV/cm$^3$, as a function of the duration of the encounter in days. We separately show the impact of a minicluster that follows a PL profile (top panel) and an NFW profile (bottom panel). We have considered an overdensity $\delta = 1$ and masses $M = 10^{-10}\,M_\odot$ (red and black lines) and $M = 10^{-12}\,M_\odot$ (blue line), with an impact parameter $b = 0.1\,\RC$ (red and blue lines) and $b = 0.5\,\RC$ (black line). We have considered AMCs which have not been perturbed. As we show in the main text, these AMC encounters with Earth would be rare.}
 	\label{fig:EarthEncounter}
\end{figure}

In Fig.~\ref{fig:EarthEncounter}, we show the enhancement of the local energy density of axions, $\rho_a/\rho_\odot$, during an encounter with an AMC, as a function of time. We fix $\rho_\odot = 0.45\,$GeV\,cm$^{-3}$ and we show time in days. The energy density $\rho_a$ is given by Eq.~\eqref{eq:PLprofile} for miniclusters with a PL profile (top panel) and by Eq.~\eqref{eq:density_NFW} for miniclusters with an NFW profile (bottom panel). We have considered an overdensity $\delta = 1$ and masses $M = 10^{-10}\,M_\odot$ (red and black lines) and $M = 10^{-12}\,M_\odot$ (blue line), with an impact parameter $b = 0.1\,\RC$ (red and blue lines) and $b = 0.5\,\RC$ (black line). The maximum enhancement is regulated solely by the ratio between the density of the AMC, which is proportional to $\rC(\delta)$ in Eq.~\eqref{eq:density_clusters}, and $\rho_\odot$, while the mass of the AMC controls the duration of the encounter. Encounters with PL profile miniclusters are shorter and have greater enhancements than those with NFW miniclusters.

The small rate and short duration of these interactions makes AMC encounters with Earth mostly irrelevant for direct axion searches. We find that an encounter has a typical duration of $\mathcal{O}(10)$ days, giving a probability of $\mathcal{O}(10^{-5})$ and $\mathcal{O}(10^{-8})$ that the Earth is currently inside such an AMC for NFW and PL AMCs respectively.  Nevertheless, the AMC fraction $f_{\rm AMC}$ and its evolution in the MW is an important quantity for direct detection. If all axions are locked up in AMCs, direct searches may be ineffective. In addition, the disruption of miniclusters can lead to streams of axions, which can produce an important contribution to the local density~\cite{Tinyakov:2015cgg, Knirck:2018knd}. We will assess these streams more carefully in future work.

\subsection{Indirect Detection}

Axion-photon conversion in astrophysical magnetic fields can lead to potentially detectable radio-wave signals. For example, axions can resonantly convert into photons in the magnetic field of a NS~\cite{Pshirkov:2007st} and non-resonantly in the Galactic center of target galaxies~\cite{Kelley:2017vaa}. The latter process is severely suppressed in favor of photon-photon pair production~\cite{Sigl:2017sew, Caputo:2018ljp, Caputo:2018vmy}. The search for these radio signals is complementary to laboratory experiments and might help disentangle the coupling $g_{a\gamma\gamma}$ from the cosmic axion abundance.

The photon production rate in these astrophysical environments depends on the local number density of axions, which is enhanced in the presence of an AMC. In Ref.~\cite{Edwards:2020afl}, we compute the expected signals from encounters between miniclusters and NSs.

\section{Discussion and Conclusion}
\label{sec:conclusion}

In this paper we quantified the degree to which tidal interactions with stars can affect the final distributions of axion miniclusters (AMCs) in the Milky Way (MW). We performed Monte Carlo simulations for AMCs on circular and eccentric orbits, and with Power-law (PL) and NFW density profiles. Importantly, we quantified the survival probability of AMCs as a function of the galactocentric radius (Fig.~\ref{fig:survival}), showing that in the inner regions of the MW, $r \lesssim \mathcal{O}(1) \,\mathrm{kpc}$, AMCs are substantially depleted. At larger radii, $r \gtrsim \mathcal{O}(10) \,\mathrm{kpc}$, AMCs have a high probability of surviving.

The interactions of AMCs with stars can also alter the properties of the surviving AMCs (Fig.~\ref{fig:histdensity}) through partial mass loss and energy injection. We have presented the procedure for deriving these properties using the results of our Monte Carlo simulations (with accompanying code and distributions at \href{https://github.com/bradkav/axion-miniclusters/}{github.com/bradkav/axion-miniclusters}~\cite{AMC_code}). This allows our results to be re-interpreted for an arbitrary initial distribution of AMCs in the MW.

As discussed in \S~\ref{sec:caveats}, for computational simplicity we made a number of assumptions. We reiterate two assumptions here as they are of primary importance for future work. Firstly, we decoupled structure formation from our tidal interactions by initially populating the MW with AMCs following a halo mass function evaluated today. Fortunately, the majority of AMCs (in particular the less massive ones) form before the formation of the MW halo, meaning that we expect only small changes to our final distributions of miniclusters if we were to follow both structure formation and tidal interactions. This concurrent evolution should be investigated more precisely in future work. Unfortunately, it may be difficult to simultaneously obtain the necessary scale and resolution needed to simulate both stars and AMCs. Secondly, we assumed a static MW halo with a fixed stellar population as measured today (see Fig.~\ref{fig:StellarDensity}). Future work should replace this assumption with a co-evolving stellar population that follows the cosmic star formation rate~\cite{Madau:2014bja} and an evolving MW halo model such as Ref.~\cite{Samland:2003ii}.

Throughout the disruption calculations, we neglected the effect of axion stars (ASs), which may form in the centers of AMCs. The formation and evolution of these ASs is still uncertain. Since we do not account for these ASs, we instead place a cut on the AMC parameter space which requires the central AS to have a smaller radius than the host AMC (referred to as `AS cut' and described in \S~\ref{sec:ASs}). For those that pass the AS cut, we neglect the potential effects of a solitonic core on the stability of the minicluster. Future work should study the response of an AMC-AS system to tidal perturbations and extend our prescription to account for this more complete description of the axion sub-structure population.

 Throughout this paper, we assumed that the entire population of AMCs in the MW had either a PL or NFW internal density profile. As discussed in \S~\ref{sec:PhaseSpace} and \S~\ref{sec:caveats}, it is still unclear which density profile best describes the overall population of miniclusters --- in fact it is likely that the high-mass miniclusters have more NFW-like profiles, while the lowest-mass miniclusters have PL-like profiles. Here, we have tried to bracket the uncertainty on the final distributions of miniclusters by using a very concentrated profile (PL) which is robust to perturbations and a more loosely bound profile (NFW) which is more easily disrupted. More work is needed to assess how the structure of these miniclusters evolves and, in particular, which AMC masses are better described by NFW or PL profiles (or perhaps some intermediate profile). By construction, our results are re-interpretable, allowing us to use this information to directly build more accurate descriptions of AMCs today when it becomes available.

In the post-inflationary scenario considered, the mass of the QCD axion has to be tuned to a specific value $\tilde m_a$ in order to reproduce the present DM abundance. If we were to set the mass of the QCD axion to be heavier than $\tilde m_a$, the axion would not be the dominant component of  DM in the Universe. This means that the predictions for the axion mass enclosed at $t_{\rm osc}$, i.e.\ the AMC mass, and the distribution of overdensities would have to be recomputed from new simulations taking into account the growth of axion overdensities around the dominant DM component. Such scenarios have never been considered in the literature, nor are they taken into account in the present work. On the other hand, an axion of mass $m_a < \tilde m_a$ is not allowed, as its present energy density would be larger than what is observed for DM. For this reason, we have chosen $m_a = \tilde m_a$ throughout the paper. Another subtlety lies in the possible range of $\tilde m_a$. We have set $\tilde m_a = 20{\rm \,\mu eV}$ following recent estimates in the literature~\cite{Klaer:2017qhr, Buschmann:2019icd}. However, uncertainties in the numerical computations hint at a range $\tilde m_a = \mathcal{O}\left(10-100\right){\rm \,\mu eV}$ where the KSVZ axion mass could lie, see Ref.~\cite{DiLuzio:2020wdo}. Possible scaling violations in the string dynamics could also suggest that a large number of axions are produced from decaying strings in the early Universe, leading to a value of the axion mass $\tilde{m}_a \approx 500 \,\mu\mathrm{eV}$ that differs from the one we have set here~\cite{Gorghetto:2020qws}. Assuming a different value of $\tilde m_a$ would lead to a different characteristic mass $M_0$, which would cause the related expressions for $M_{\rm min}$ and $M_{\rm max}$ to be modified. However the analysis we have presented could be straightforwardly re-applied --- as we have stressed --- and the results would not change qualitatively.

Previous work has considered the effects of tidal interactions on AMCs~\cite{Tinyakov:2015cgg,2017JETP..125..434D}. In particular, Ref.~\cite{2017JETP..125..434D} considered AMCs with PL profiles $\propto r^{-1.8}$ and found that around 2-5\% are destroyed at the Solar position. At the same position, we find that >99\% of AMCs with a PL profile survive, while those with NFW profiles are more easily stripped with a survival probability $\sim 94\%$ (see Fig.~\ref{fig:survival} without AS cut). This difference can easily be explained by the fact that our PL profile is significantly steeper than the one considered in Ref.~\cite{2017JETP..125..434D}, making our AMCs more robust to perturbations from stellar encounters. We also build upon Ref.~\cite{2017JETP..125..434D} in two key ways: firstly, we extend the calculation of the properties of AMCs beyond the Solar position to the entire Galaxy; and secondly, we account for the injection of energy from non-fatal interactions, allowing us to more realistically describe disruption probabilities and the properties of AMCs today.

\smallskip

In our companion paper~\cite{Edwards:2020afl}, we use the results of our Monte Carlo simulations to predict indirect signals from AMCs interacting with neutron stars (NSs). More precisely we: {\bf i)} simulate the expected encounter rate for NSs passing through AMCs; and {\bf ii)} estimate the expected radio signal from the conversion of axions within the minicluster into photons as they interact with the NS magnetosphere. These results are complementary to the continuous radio emission that is expected to come from axions in the Galactic halo interacting with NSs~\cite{Pshirkov:2007st, Huang:2018lxq, Hook:2018iia, Safdi:2018oeu}.

As we showed in Sec.~\ref{sec:apps}, correctly calculating the AMC population today can have significant implications for a variety of observational channels. It is therefore of upmost importance to understand the interactions of these structures with their environment. Our results represent a fundamental step towards characterizing these interactions within the MW.

\begin{acknowledgements}

The authors would like to thank Ciaran O'Hare for a careful reading of the manuscript. We thank Sebastian Baum, Gianfranco Bertone, Malte Buschmann, Matthew Lawson, David J. E. Marsh, M.C. David Marsh, Alexander Millar, Lina Necib, Javier Redondo, and Ben Safdi for providing insightful comments. T.E. acknowledges support by the Vetenskapsr{\aa}det (Swedish Research Council) through contract No.  638-2013-8993 and the Oskar Klein Centre for Cosmoparticle Physics. T.E was also supported in part by the research environment grant ‘Detecting Axion Dark Matter In The Sky And In The Lab (AxionDM)’ funded by the Swedish Research Council (VR) under Dnr 2019-02337. T.E. and C.W. are supported by the NWO through the VIDI research program ``Probing the Genesis of Dark Matter'' (680-47-5). L.V. is supported through the research program ``The Hidden Universe of Weakly Interacting Particles'' with project number 680.92.18.03 (NWO Vrije Programma), which is partly financed by the Nederlandse Organisatie voor Wetenschappelijk Onderzoek (Dutch Research Council), and acknowledges support from the European Union's Horizon 2020 research and innovation programme under the Marie Sk{\l}odowska-Curie grant agreement No.~754496 (H2020-MSCA-COFUND-2016 FELLINI). B.J.K. thanks the Spanish Agencia Estatal de Investigaci\'on (AEI, MICIU) for the support to the Unidad de Excelencia Mar\'ia de Maeztu Instituto de F\'isica de Cantabria, ref. MDM-2017-0765.

Some of this work was carried out on the Dutch national e-infrastructure with the support of SURF Cooperative. Finally, we acknowledge the use of the Python scientific computing packages NumPy~\cite{numpy,Harris:2020xlr} and SciPy~\cite{scipy}, as well as the graphics environment Matplotlib~\cite{Hunter:2007}.

\end{acknowledgements}

\bibliography{AxionBib}

\newpage
\onecolumngrid

\appendix

\section{Tidal stripping from the Milky Way host halo}
\label{app:host_halo}

Here, we quantify the impact of tidal stripping of AMCs by the Milky Way host halo. For a point mass $M_\mathrm{AMC}$ on a circular orbit at a galactocentric radius $r$, the tidal radius can be written as~\cite{vandenBosch:2017ynq}:
\begin{equation}
    R_\mathrm{t}=r\left[\frac{M_\mathrm{AMC} / M_\mathrm{MW}(r)}{3-\left.\frac{\mathrm{d} \ln M_\mathrm{MW}}{\mathrm{~d} \ln r}\right|_{r}}\right]^{1 / 3}\,,
\end{equation}
where $M_\mathrm{MW}(r)$ is the mass of the Milky Way DM halo enclosed within a radius $r$. At the tidal radius, the tidal forces due to the host halo exceed the self-gravity from the orbiting point mass. If the AMC radius exceeds the tidal radius, then the AMC is likely to undergo tidal stripping by the host halo. 

\begin{figure}[ht!]
 	\includegraphics[width=0.5\linewidth]{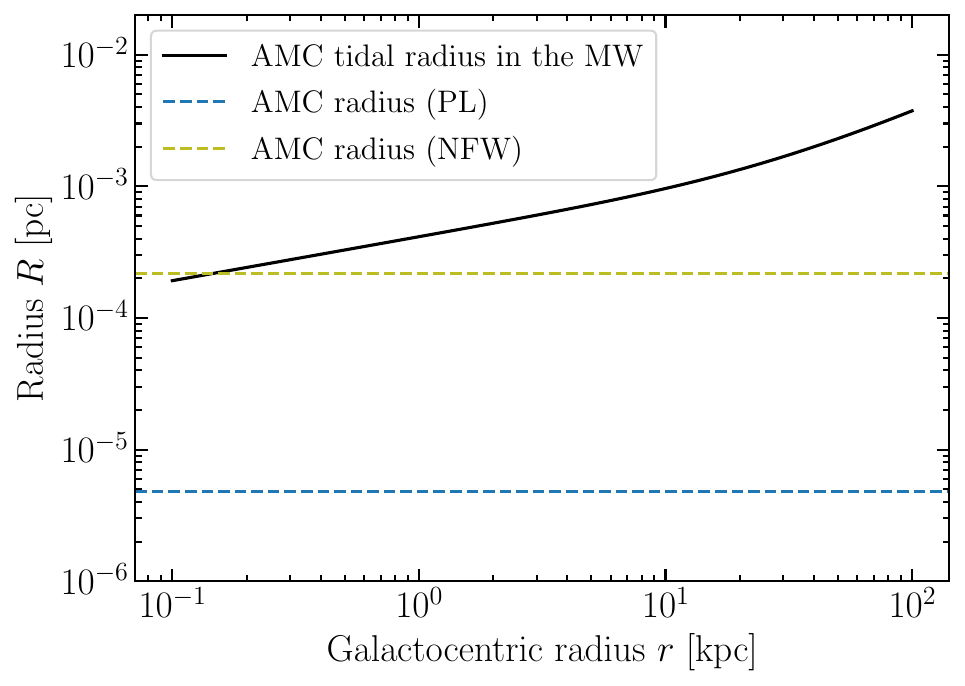}
 	\caption{Tidal radius $R_t$ for AMCs in the Milky Way. We also show the AMC radii assumed in the main text for PL (blue dashed) and NFW (olive dashed) AMC density profiles, fixing $M_\mathrm{AMC} = 10^{-10}\,M_\odot$ and $\delta = 1.0$. }
 	\label{fig:TidalRadii} 
\end{figure}

In Fig.~\ref{fig:TidalRadii}, we compare this tidal radius (solid black) with our assumed AMC radius for PL (dashed blue) and NFW (dashed olive) AMCs, fixing $M_\mathrm{AMC} = 10^{-10}\,M_\odot$ and $\delta = 1.0$. Both the tidal radius and AMC radius scale in the same way with $M_\mathrm{AMC}^{1/3}$, so this comparison extends straightforwardly to other AMC masses. For larger values of the overdensity $\delta$, the AMC radius would be smaller, making such AMCs more robust to tidal stripping. This means that the results we show in Fig.~\ref{fig:TidalRadii} for $\delta = 1$ are conservative.

For PL profiles, we see that the tidal radius is significantly larger than the size of the AMC, even down to small Galactocentric radii, meaning that PL AMCs should be robust to tidal stripping. We therefore ignore any tidal stripping by the MW host halo in the case of PL AMCs. For NFW profiles, instead, the tidal radius may be comparable to the size of the AMC, particularly in the inner Galaxy. These estimates were calculated assuming a concentration of $c = 100$, while AMCs are expected to have a much larger concentration ($c \sim 10^4$) before they are accreted into the MW. The picture we consider then is that  initially large AMCs were accreted and then tidally stripped by the MW halo, down to a concentration of $c \sim 100$, at which point their physical radius becomes comparable to the tidal radius. Mass is lost by the AMCs in this process, meaning that the $c \sim 100$ AMCs we consider should have a smaller mass than objects drawn from the \textit{initial} HMF described in \S~\ref{sec:HMF}.

Analytic arguments and numerical simulations of DM substructure suggest that the rate of mass loss due to tidal stripping is suppressed by $(m/M)^{\zeta}$, for a substructure of mass $m$ and host halo of mass $M$~\cite{vandenBosch:2004zs}. More carefully, we can write \cite[Sec.~2.2]{Jiang:2014nsa}:
\begin{equation}
\label{eq:mdot}
    \dot{m}=-\mathcal{A} \frac{m}{\tau_{\mathrm{dyn}}}\left(\frac{m}{M}\right)^{\zeta}\,,
\end{equation}
where $\tau_{\mathrm{dyn}}$ is the dynamical time of the MW halo (approximately $2.4\,$Gyr today).\footnote{Note that Ref.~\cite{Hiroshima:2018kfv} demonstrated the validity of Eq.~\eqref{eq:mdot} for the mass ratios considered in this work.} Fixing $\mathcal{A} = 1.34$ and $\zeta = 0.07$~\cite{Jiang:2014nsa} and assuming a static MW, we can solve Eq.~\eqref{eq:mdot} to obtain an estimate for the fractional mass-loss over the age of the Galaxy. Given an initial AMC mass of $M_i$, we find that the mass after stripping is:
\begin{equation}
    M_\mathrm{stripped} = M_i \left[1 + \zeta \left(\frac{M_i}{M_\mathrm{MW}}\right)^\zeta \left(\frac{\mathcal{A} t_\mathrm{MW}}{t_\mathrm{dyn}}\right)\right]^{-1/\zeta}\,,
\label{eq:M_stripped}
\end{equation}
where $t_\mathrm{MW} = 13.5 \times 10^{9}\,\mathrm{yr}$ and $M_\mathrm{MW} = 10^{12}\,M_\odot$.

\begin{figure}[ht!]
 	\includegraphics[width=0.5\linewidth]{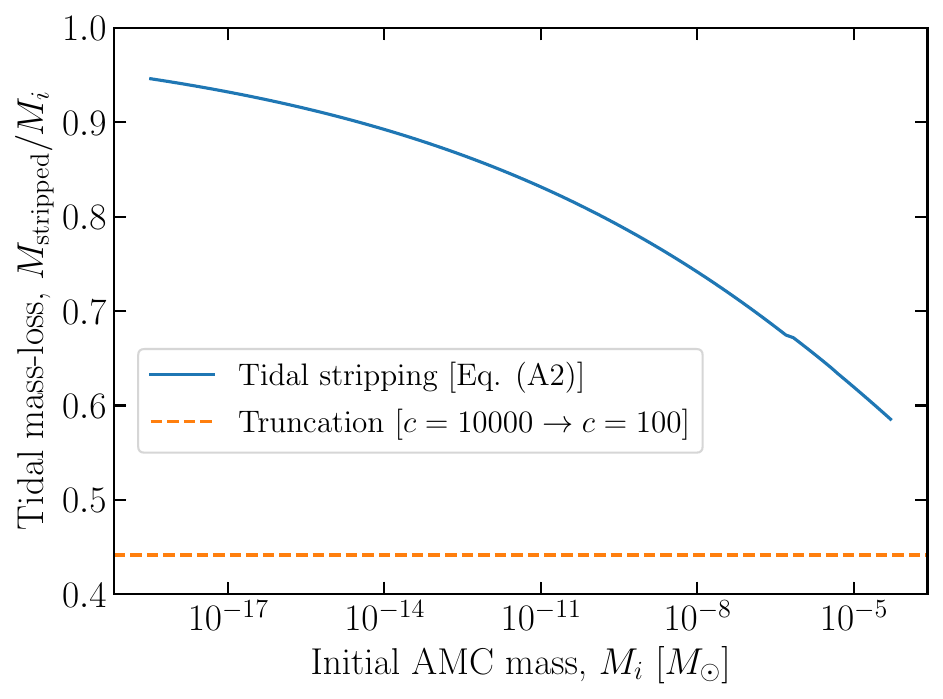}
 	\caption{Mass of NFW-profile AMCs after tidal stripping by the host halo $M_\mathrm{stripped}$ (solid blue). For NFW AMCs, we apply this mass-loss fraction as a correction to the initial mass function in the main text. For comparison, we also show the mass-loss expected by stripping an NFW halo from $c = 10^4$ to $c = 100$.}
 	\label{fig:TidalMassLoss} 
\end{figure}

The AMC mass after tidal stripping due to the host halo is illustrated in Fig.~\ref{fig:TidalMassLoss}. We see that NFW AMCs should lose 5-40\% of their initial mass in this way. For comparison, we show also the mass-loss which would be estimated by starting with an NFW density profile with concentration $c = 10^4$ (as expected for field AMCs at $z = 0$) and truncating the profile at $R = R_\mathrm{AMC}/100$ (leading effectively to a profile with $c = 100$, as we assume in the main text). This leads to a mass-loss of $f_\mathrm{NFW}(100)/f_\mathrm{NFW}(10^4) \sim 50\%$ (orange dashed line), in rough agreement with the results of Eq.~\ref{eq:M_stripped} for the heaviest AMCs. 

We incorporate this tidal stripping of NFW AMCs by correcting the initial HMF using Eq.~\eqref{eq:M_stripped}:
\begin{equation}
    P(M_\mathrm{stripped}) = P(M_i) \frac{\mathrm{d}M_i}{\mathrm{d}M_\mathrm{stripped}}\,.
\end{equation}
We then use this distribution $P(M_\mathrm{stripped})$ as the \textit{initial} distribution $P_i(M)$ described in \S~\ref{sec:PhysicalProperties} of the main text. This amounts to assuming that the tidal stripping from the MW halo happens first, followed by perturbations from stars. Of course, these effects will occur concurrently over the age of the MW. However, as we have demonstrated, the disruption of AMCs due to stellar encounters is largely independent of AMC mass (depending instead predominantly on the mean AMC density), so we do not expect a mass-loss of $\mathcal{O}(10 - 40)\%$ due to the host halo to substantially impact the survival probability of AMCs.

\section{Stellar Population in the Milky Way}
\label{sec:stellar_population}

\begin{figure}[ht!]
 	\includegraphics[width=0.7\linewidth]{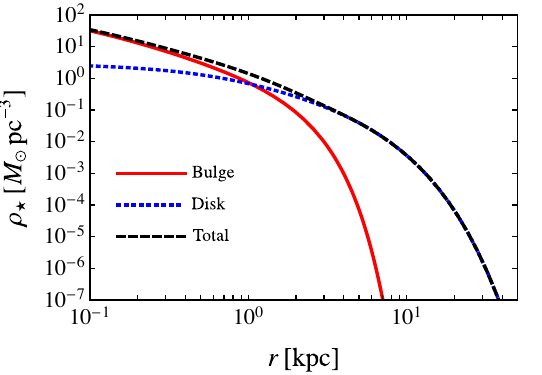}
 	\caption{Stellar density used in the Monte Carlo simulations as a function of the galactocentric radius $r$, averaged over the cylindrical Galactic height. We show the density of stars in the bulge (red line), in the disk (blue line), 
 	and the sum of the two (black line).}
 	\label{fig:StellarDensity} 
\end{figure}

We assume that the Galactic distribution of stars can be decomposed into an axially-symmetric bulge in the innermost region and an axially-symmetric stellar disk.
We model these distributions in terms of the galactocentric cylindrical coordinates $r_\mathrm{cyl}$ and $z_\mathrm{cyl}$, which describe the radial distance from the axis of symmetry and the height from the Galactic plane respectively. In more detail:
\begin{enumerate}
    \item The essential features of the stellar bulge can be captured by fitting the stellar density profile by a truncated Power-law distribution~\cite{Binney:1996sv, Bissantz:2001wx}
    \begin{equation}
        \label{eq:stellar_bulge}
        \rho_{\star}^{\rm bulge}(r_\mathrm{cyl}, z_\mathrm{cyl}) = \rho_0^{\rm bulge}\frac{e^{-\(r'/r_{\rm cut}\)^2}}{\(1 + r'/r_0\)^\lambda}\,,
    \end{equation}
    where we fix the parameters according to Ref.~\cite{McMillan:2011wd}, namely the core density $\rho_0^{\rm bulge} \approx 99.3\,M_\odot/$pc$^3$, the radius $r' = \sqrt{r_\mathrm{cyl}^2 + (z_\mathrm{cyl}/q)^2}$ with $q = 0.5$, the bulge cutoff $r_0 = 0.075{\rm \,kpc}$, 
    the exponent $\lambda = 1.8$, and $r_{\rm cut} = 2.1\,$kpc. Note that different models exist in the literature that provide fits which deviate from an axially symmetric solution, for example including triaxality, see Refs.~\cite{Binney:1996sv, 2009A&A...498...95V, 2005A&A...439..107L, 2012A&A...538A.106R}.
    \item The stellar disk component in the Milky Way (MW) can be further subdivided into a thin (t) and a thick (T) disk, as inferred by the different chemical compositions and spatial distributions~\cite{Bensby:2004pc, Juric:2005zr}, with the thin disk hosting younger stars which are more concentrated around the Galactic plane with respect to the thick disk counterpart. Both disk distributions are commonly described by a double exponential model~\cite{Bahcall:1980fb, 1998MNRAS.294..429D},
    \begin{equation}
        \label{eq:stellar_disc}
        \rho_{\star}^{\rm disk}(r_\mathrm{cyl}, z_\mathrm{cyl}) = \frac{\Sigma_{\star}}{2H}\,\exp\(-\frac{r_\mathrm{cyl}}{L} - \frac{|z_\mathrm{cyl}|}{H}\)\,,
    \end{equation}
    where $H$ and $L$ are the scale height and the scale length, respectively, and $\Sigma_{\star}$ is the surface stellar density.
    We fix the parameters as in the best-fitting model presented in Table~2 of Ref.~\cite{McMillan:2011wd}, namely the surface densities $\Sigma_{\star}^{\rm t} = 816.6\,M_\odot\,{\rm pc}^{-2}$ and $\Sigma_{\star}^{\rm T} = 209.5\,M_\odot\,{\rm pc}^{-2}$, together with the scale lengths $L^{\rm t} = 2.90\,$kpc and $L^{\rm T} = 3.31\,$kpc. Similarly to Ref.~\cite{McMillan:2011wd}, we use the bias-corrected values of the scale heights from Table~10 of Ref.~\cite{ Juric:2005zr}, namely $H^{\rm t} = 0.3\,$kpc and $H^{\rm T} = 0.9\,$kpc. These values are consistent with other independent analyses~\cite{Siegel:2002vr}.
\end{enumerate}
In principle, a halo component dominates the stellar density in the outskirts of the MW, see for example Ref.~\cite{Juric:2005zr}. However, using the parametrization in Ref.~\cite{Juric:2005zr}, we find that the halo component only becomes important for radii $r\gtrsim 30\,$kpc at which point the effect of tidal stripping from stars is negligible. In addition, we were unable to find a consistent model that simultaneously fit halo, bulge, and disk components to the MW stellar population. For these reasons, we have not included this additional component.
In our work, we then define the total stellar energy density profile as $\rho_{\star} = \rho_{\star}^{\rm bulge} + \rho_{\star}^{\rm disk}$.
Figure~\ref{fig:StellarDensity} shows the stellar distribution used for this paper 
averaged over the cylindrical Galactic height $z_\mathrm{cyl}$, as a function of the galactocentric radius $r$. 
We show the density of stars in the bulge Eq.~\eqref{eq:stellar_bulge} (red line), in the disk Eq.~\eqref{eq:stellar_disc} (blue line), 
and the sum of the two (black line).

\section{Overdensity Expressions}
\label{sec:overdensity}

We report the expression used in this work for the distribution of overdensities, which we based on the results in Ref.~\cite{Buschmann:2019icd}:

\begin{equation}
	\frac{\mathrm{d}f_{\rm AMC}}{\mathrm{d}\delta} = \frac{P_\delta\,A}{1 + (\delta/\delta_F)^S}\,,
	\label{eq:dfddelta}
\end{equation}
where $S = 4.7$, $A = 1/2.045304$, and $\delta_F = 3.4$. We also set
\begin{equation}
    P_\delta =
    \begin{cases}
        \exp\[-\(\frac{x^2}{2\sigma^2}\)^{d/2}\]\,, & x \leq \sigma\,\alpha_d\,,\\
        B_1\,\(\frac{\sigma\,C + x}{\sigma\,B_2}\)^{-n}\,, & x > \sigma\,\alpha_d\,,\\
    \end{cases}
\end{equation}
where $x = \ln\(\delta/\delta_G\)$ with $\delta_G = 1.06$, and where $d = 1.93$, $\alpha_d = -0.21$, $n = 11.5$, $\sigma = 0.448$, and
\begin{eqnarray}
    B_1 &=& \exp\[-\(\frac{\alpha_d^2}{2}\)^{d/2}\]\,,\\
    B_2 &=& \(\frac{2}{\alpha^2}\)^{d/2}\,\frac{n\,|\alpha_d|}{d}\,,\\
    C &=& B_2 + |\alpha_d|\,.
\end{eqnarray}
Note that we use a slightly amended functional form compared to Ref.~\cite{Buschmann:2019icd}, in order to ensure that the function is smooth and continuous. 

\section{AMC Distribution Functions}
\label{app:distributionfunctions}

Here, we elaborate on the evaluation of the initial AMC distribution functions for the Power-law (PL) and NFW internal density profiles. Assuming isotropic and spherically symmetric orbits, we begin with the Eddington Inversion formula~\cite[p. 290]{BinneyTremain:2008}:
\begin{equation}
\label{eq:Eddington}
    f(\mathcal{E}) \equiv \frac{1}{\sqrt{8}\pi^2} \int_{0}^{\mathcal{E}} \frac{1}{\sqrt{\mathcal{E} - \Psi}}\frac{\mathrm{d}^2\rho}{\mathrm{d}\Psi^2}\,\mathrm{d}\Psi\,.
\end{equation}
It is cumbersome to re-compute the distribution function for different choices of the AMC mass $M_\mathrm{AMC}$ and radius $R_\mathrm{AMC}$, so it will be more useful to work in terms of the dimensionless energy, potential, and density: $\epsilon = \mathcal{E}/\Psi_0$,  $\psi = \Psi/\Psi_0$, and $\varrho = \rho/\rho_\mathrm{AMC}$ respectively, with $\Psi_0 \equiv G M_\mathrm{AMC}/R_\mathrm{AMC}$.\footnote{Recall here that $\rho_\mathrm{AMC}$ is the characteristic AMC density. In the case of PL profiles, this is equal to the mean density  $\rho_\mathrm{AMC} = \bar{\rho}$, while for NFW profiles $\rho_\mathrm{AMC} = \bar{\rho} \,c^3/\left(3 f_\mathrm{NFW}(c)\right)$, with $c \approx 100$ (see Eq.~\eqref{eq:NFWscaleradius} and surrounding text).} With these definitions, Eq.~\eqref{eq:Eddington} becomes:
\begin{equation}
    f(\mathcal{E}) =  \frac{\rho_\mathrm{AMC}} {\Psi_0{}^{3/2}} \hat{f}\left( \mathcal{E}/\Psi_0\right)\,; \qquad \text{with}\quad \hat{f}(\epsilon) = \frac{1}{\sqrt{8}\pi^2}\int_{0}^{\epsilon} \frac{1}{\sqrt{\epsilon -  \psi}}\frac{\mathrm{d}^2\varrho}{\mathrm{d}\psi^2}\,\mathrm{d}\psi\,.
\end{equation}

For the power-paw profile, defining $x = R/R_\mathrm{AMC}$, we have:
\begin{align}
    \begin{split}
        \varrho_\mathrm{PL}(x) &= \begin{cases}
            \frac{1}{4} x^{-9/4}& \qquad  \text{ for } x < 1\\
            0& \qquad  \text{ for } x > 1
        \end{cases}\\
        \psi_\mathrm{PL}(x) &=  
\begin{cases}
1+ 4  (x^{-1/4} - 1)& \qquad \text{ for } x < 1\\
x^{-1} &\qquad \text{ for } x > 1
\end{cases}\,.
    \end{split}
\end{align}
For the NFW profile, 
\begin{align}
    \begin{split}
        \varrho_\mathrm{NFW}(x) &= 
        \begin{cases}
            \frac{1}{c\,x\,(1 + c \,x)^2}& \qquad  \text{ for } x < 1\\
            0& \qquad  \text{ for } x > 1
        \end{cases}\,\\
        \psi_\mathrm{NFW}(x) &= 
        \begin{cases}
1 + \frac{c}{f_\mathrm{NFW}(c)} \left(\frac{\ln(1 + x)}{x} - \ln 2\right)& \qquad \text{ for } x < 1\\
x^{-1} &\qquad \text{ for } x > 1
\end{cases}\,,
    \end{split}
\end{align}
where $c = 100$ is the assumed truncation parameter and $f_\mathrm{NFW}(c) = \ln(1 + c) - c/(1+c)$. In both cases, we have implemented a hard truncation of the density profile by setting $\varrho(x > 1) = 0$. With these expressions, the density cannot be written analytically in terms of the potential, so the integral in $\hat{f}(\epsilon)$ must be evaluated numerically for both PL and NFW profiles. 

Using these forms for the distribution function, we can estimate the mass-loss in a given encounter. We can also estimate the velocity dispersion (as a function of radius) of the AMC as:
\begin{equation}
    \sigma^2(x) = \frac{\Psi_0}{\varrho(x)} 4 \pi  \int_0^{\psi(x)} \left[2 (\psi(x) - \epsilon)  \right]^{3/2} \hat{f}(\epsilon)\,\mathrm{d}\epsilon\,,
\end{equation}
from which we obtain the mean velocity dispersion of the AMC, which we evaluate numerically.

The gravitational binding energy is calculated as:
\begin{align}
\label{eq:Ebind_app}
    E_\mathrm{bind} = \int_0^{R_\mathrm{AMC}} \frac{G M_<(R)}{R}\times 4 \pi R^2 \rho(R) \,\mathrm{d}R\,,
\end{align}
with the enclosed mass
\begin{align}
\label{eq:Menc_app}
    M_<(R) = \int_0^R 4 \pi R^2 \rho(R)\,\mathrm{d}R\,.
\end{align}
Immediately after the stellar interaction, we can calculate the final density profile $\rho_f(R)$ by assuming that unbound particles are instantaneously removed to infinity. The change in density due to the removal of these particles is:
\begin{align}
    \Delta\rho(R) = 4 \pi \int_0^{\mathrm{min}[\Delta\mathcal{E}(R), \Psi(R)]} \,\sqrt{2 (\Psi(R) - \mathcal{E})} f(\mathcal{E}) \,\mathrm{d}\mathcal{E}\,,
\end{align}
such that:
\begin{align}
    \rho_f(R) = \rho_i(R) - \Delta \rho(R)\,,
\end{align}
for initial density $\rho_i$. The binding energy of the AMC immediately after the interaction can then be calculated by substituting $\rho_f$ into Eqs~\eqref{eq:Ebind_app} and \eqref{eq:Menc_app}. 

Given an AMC density distribution $\rho(r)$, we have defined the quantities
\begin{eqnarray}
    \alpha^2 &=& \frac{4\pi}{M_{\rm AMC}\,R_{\rm AMC}^2}\int_0^{R_{\rm AMC}}\mathrm{d}r r^4\rho(r)\,,\\
    \beta &=& \frac{4\pi\,R_{\rm AMC}}{{M_{\rm AMC}}^2}\int_0^{R_{\rm AMC}}\mathrm{d}r r\rho(r)M_{\rm enc}(r)\,,\\
\end{eqnarray}
where $M_{\rm AMC}$ is the mass of the AMC, $R_{\rm AMC}$ the truncation radius, and $M_{\rm enc}(r)$ the mass enclosed in the distribution within the radius $r$. More specifically, for the distributions we have adopted in this paper, we find
\begin{equation}
    \alpha^2 =
    \begin{cases}
        \frac{3}{c^2} + \frac{1}{2 f_{\rm NFW}(c)}\,\frac{c - 3}{c + 1}\, &\quad  \hbox{(NFW)}\,,\\
        \frac{3}{11}\,& \quad \hbox{(PL)}\,.
    \end{cases}
\end{equation}
and
\begin{equation}
    \beta =
    \begin{cases}
        \frac{c^3 - 2 c (1 + c) f_{\rm NFW}(c)}{2 (1 + c)^2 f_{\rm NFW}(c)^2}\, &\quad  \hbox{(NFW)}\,,\\
        \frac{3}{2}\,& \quad \hbox{(PL)}\,,
    \end{cases}
\end{equation}
where the function $f_{\rm NFW}(c)$ has been defined after Eq.~\eqref{eq:NFWscaleradius}. For a detailed study of the NFW case, see also Ref.~\cite{Mo:1997vb}.

\section{Comparing NFW internal density profiles}
\label{app:NFWassumptions}

\begin{figure}[tb!]
\includegraphics[width=0.5\textwidth]{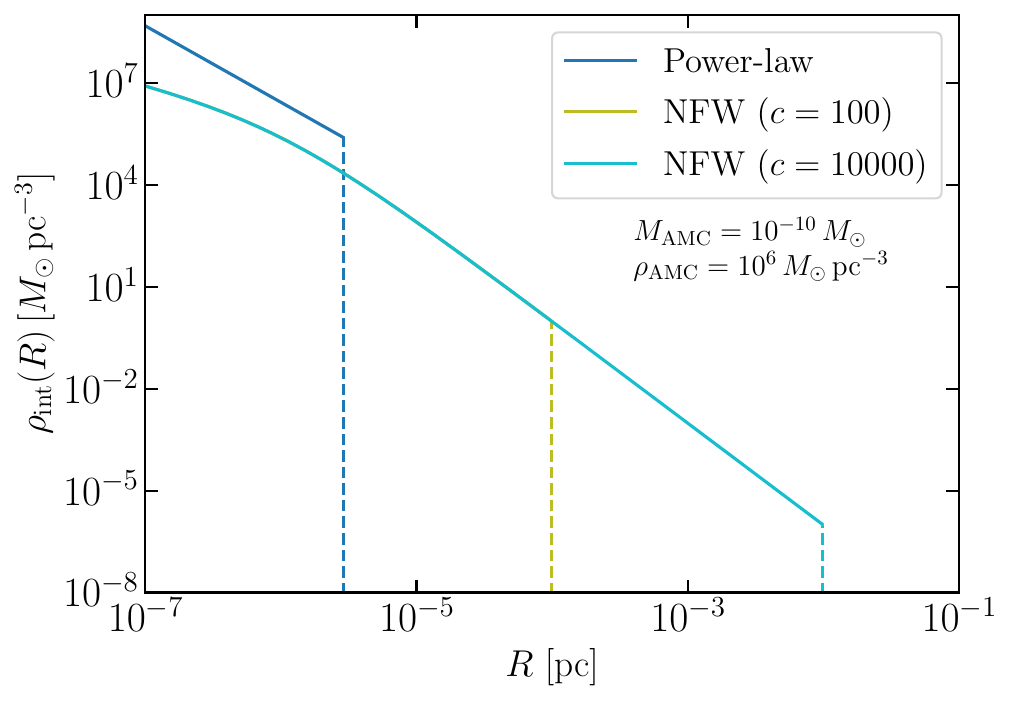}
	\caption{Examples of AMC density profiles.  The NFW profile with truncation parameter $c = R_\mathrm{AMC}/r_s = 100$ assumed in the main text can be compared with the more diffuse profile with $c = 10^4$ which we also consider here. In the case of $c = 100$, we assume a total mass of $M_\mathrm{AMC} = 0.44 \times 10^{-10}\,M_\odot$, to give the same central density as the $c = 10^4$ profile.}
	\label{fig:NFW_profiles}
\end{figure}

In the main text, we make an identification between then AMC properties ($M_\mathrm{AMC}$, $\delta$) and the corresponding internal NFW profile ($\rho_s$, $r_s$) by assuming a fixed truncation parameter $c = R_\mathrm{AMC}/r_s = 100$. In this appendix, we explore the impact of varying the truncation parameter $c$. Specifically, we consider also the case of an NFW profile with a truncation parameter $c = 10^4$, typical of the concentration of isolated AMCs at $z=0$~\cite{Ellis:2020gtq}. Figure~\ref{fig:NFW_profiles} shows a comparison of NFW profiles with truncation parameters of $c = 100$ and $c = 10^4$, for a particular choice of  characteristic density $\rho_\mathrm{AMC} = 10^6 \,M_\odot \,\mathrm{pc}^{-3}$. For the PL and $c = 10^4$ NFW profiles, we assume an AMC mass of $M_\mathrm{AMC} = 10^{-10}\,M_\odot$. For $c = 100$, we assume $M_\mathrm{AMC} = 0.44 \times 10^{-10}\,M_\odot$, which leads to the same central density as the $c = 10^4$ model. This reflects the physical motivation for the $c = 100$ model: a diffuse AMC is likely to be tidally stripped by the MW halo (as described in Appendix~\ref{app:host_halo}) resulting in a mass loss of $\mathcal{O}(50\%)$ and leaving behind a more compact AMC.

We have repeated the procedure described in Appendix~\ref{app:distributionfunctions} to determine the distribution function and properties of this more diffuse $c=10^4$ NFW profile. For the case of $c = 10^4$, we find a mean squared radius of $\left\langle R^{2}\right\rangle=\alpha^{2} R_{\mathrm{AMC}}^{2}$ with $\alpha^2 = 0.061$ and a binding energy of $E_\mathrm{bind} = \beta G M_\mathrm{AMC}{}^2/R_\mathrm{AMC}$ with $\beta = 74.0$. For a given stellar encounter, the injected energy scales as
\begin{equation}
    \frac{\Delta E}{E_\mathrm{bind}} \propto \frac{\alpha^2}{\beta}\frac{1}{\bar{\rho}}\,,
\end{equation}
where $\bar{\rho}$ is the mean density of the AMC. The ratio $\alpha^2/\beta$ is roughly 200 times smaller for NFW profiles with $c= 10^4$ than those with $c=100$. However, the mean density for $c = 10^4$ is $\sim 10^6$ times smaller than for $c = 100$, meaning that a given stellar encounter injects a much larger amount of energy (as a fraction of $E_\mathrm{bind}$) into an AMC with $ c= 10^4$. We might therefore expect that miniclusters with larger values of $c$ will be more easily disrupted in the Milky Way.

\begin{figure}[tb!]
\includegraphics[width=0.5\textwidth]{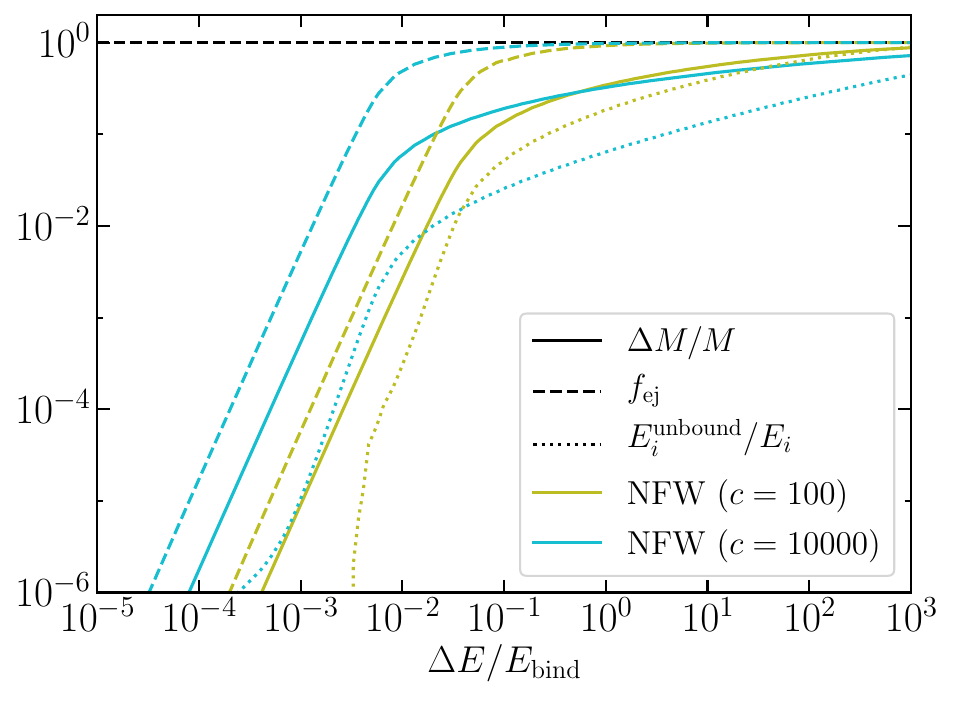}
	\caption{Response of NFW miniclusters with different truncation parameters $c$ to stellar perturbations, as a function of the injected energy $\Delta E$. We plot the fractional mass loss (solid lines), the fraction of injected energy carried away by ejected particles (dashed lines) and the fraction of the initial AMC energy in particles which will eventually be unbound (dotted lines). See \S~\ref{sec:Perturbation} for more details.}
	\label{fig:NFW_massloss}
\end{figure}

Following the discussion in Sec.~\ref{sec:stripping}, we can then determine the response of $c= 10^4$ miniclusters to perturbations. In Fig.~\ref{fig:NFW_massloss} we show the fraction of mass lost from the minicluster (solid lines), the fraction of injected energy carried away by ejected particles (dashed lines), and the fraction of the initial minicluster energy stored in particles which will eventually become unbound (dotted lines). At small values of $\Delta E/E_\mathrm{bind}$, we find that a larger fraction of mass is lost from AMCs with $c = 10^4$ compared to $c = 100$; this is because in the former case particles in the diffuse outskirts of the minicluster can be more easily unbound. However, we also find that the fraction of energy carried away by unbound particles $f_\mathrm{ej}$ is always larger for $c = 10^4$. As a result, we expect that the `remnant' AMC left behind after the stellar interaction will typically be more dense than before the interaction. Similar behaviour was described in the main text for $c=100$ NFW profiles, though the effect should be even more pronounced here for $c=10^4$ profiles.

In order to explore the behaviour of the AMCs under repeated perturbations, we consider a toy setup. We generate two sets of $10^5$ AMCs with NFW internal density profiles, one set with ($c = 10^4$, $M_i = 10^{-10}\,M_\odot$) and the other with ($c = 100$, $M_i =0.44 \times 10^{-10}\,M_\odot)$. In all cases, we fix the initial overdensity parameter to be $\delta = 1.55$ ($\rho_\mathrm{AMC} = 10^{6}\,M_\odot \,\mathrm{pc}^{-3}$). We then apply the Monte Carlo procedure described in Sec.~\ref{sec:Simulations} to evolve the AMCs to today, assuming circular orbits at a galactocentric radius of $r_\mathrm{GC} = 8\,\mathrm{kpc}$. In this case, we do not include the correction due to tidal stripping from the MW halo, as described in Appendix~\ref{app:host_halo}. 

In Fig.~\ref{fig:NFW_comparison}, we show the final distributions of AMC masses $M_\mathrm{AMC}$, radii $R_\mathrm{AMC}$ and mean internal density $\bar{\rho}$. Dashed vertical lines mark the initial values of each property at the start of the simulations. We find that the typical final mass of the $c = 10^4$ miniclusters is smaller than for $c =100$ by a factor $\mathcal{O}(3)$, as more mass is stripped away from these more diffuse objects. However, we find that the final radius and final mean density of the AMCs is similar in the two cases. This is particularly striking in the case of $\bar{\rho}$, for which the $c = 10^4$ AMCs begin the simulations with a mean density which is a factor of $\sim10^6$ smaller. As described above, the stellar interactions which may efficiently strip mass from the outskirts of the $c = 10^4$ miniclusters do not inject large amounts of energy into the remnant. These interactions therefore substantially increase the mean AMC density. It therefore appears that NFWs with different concentrations are likely to be stripped to leave behind remnants of similar densities.

Finally, extrapolating these results to the full AMC mass function in Eq.~\eqref{eq:mcdistribution} (but still keeping a single fixed initial density), we can calculate the fraction of AMCs which would survive and also pass the axion star (AS) cut. We find very similar survival probabilities for the two concentrations, $p_\mathrm{surv}^{c = 100} = 9.7 \times 10^{-3}$ and $p_\mathrm{surv}^{c = 10^4} = 1.0 \times 10^{-2}$ (where we have not factorized out the fraction of AMCs which initially pass the AS cut). We therefore conclude that the initial choice of the truncation parameter for the NFW density profile does not strongly influence the survival probability or the final properties of the perturbed miniclusters, save for a correction factor of $\mathcal{O}(3)$ to the final AMC mass. This difference is at least partially compensated for in the main analysis by the initial mass loss of $5-40\%$ which we apply to NFW profiles, coming from tidal stripping due to the host halo of the MW (Appendix~\ref{app:host_halo}).

\begin{figure*}[tb!]
\includegraphics[width=1.\textwidth]{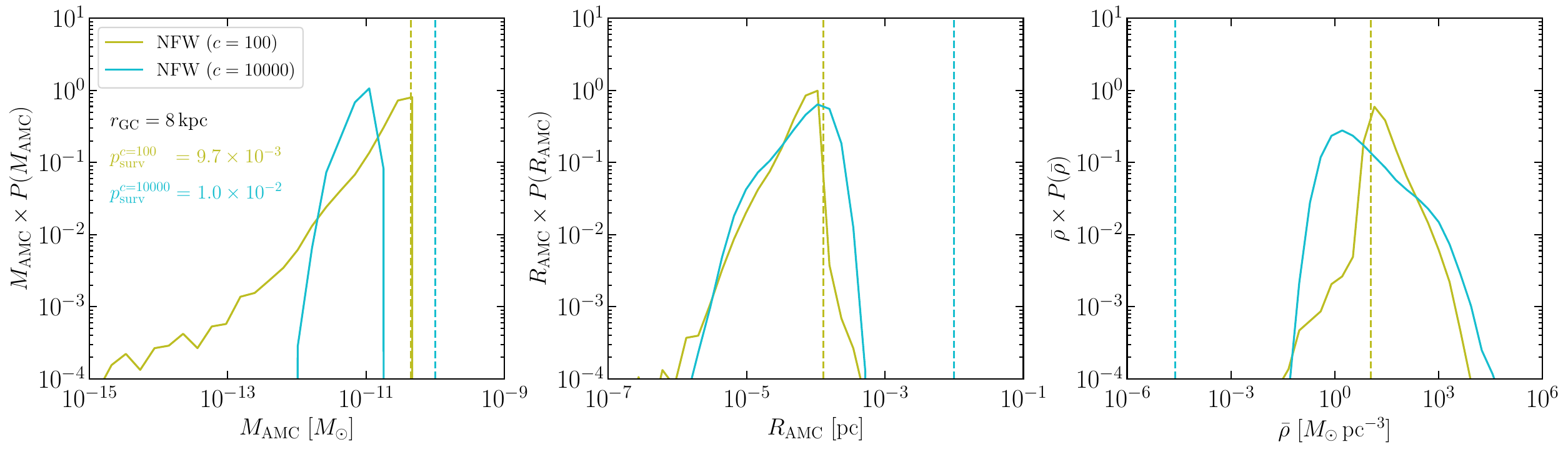}
	\caption{Comparison of final AMC properties for NFW profiles with different truncation parameters $c$. We assume an initial mass of $M_i = 10^{-10}\,M_\odot$ ($M_i = 0.44 \times 10^{-10}\,M_\odot$) for the $c = 10^4$ ($c=100$) profile and an initial characteristic density of $\rho_\mathrm{AMC} = 10^{6}\,M_\odot \,\mathrm{pc}^{-3}$. We show the final probability distributions of the mass $M_\mathrm{AMC}$, radius $R_\mathrm{AMC}$, and  mean density $\bar{\rho}$ (with the initial values shown as vertical dashed lines). In this toy sample, we assume AMCs on circular orbits at a galactocentric radius of $8\,\mathrm{kpc}$. We list the fraction $p_\mathrm{surv}$ of AMCs which both survive and pass the AS cut (extrapolating to the full mass function), though we have not applied an AS cut to the final distributions.}
	\label{fig:NFW_comparison}
\end{figure*}

\end{document}